\documentclass{jfm}
\usepackage{graphicx}
\usepackage{gensymb}
\usepackage{nicematrix}
\usepackage{newtxtext}
\usepackage{newtxmath}
\usepackage{natbib}
\usepackage{hyperref}
\usepackage{wrapfig}
\usepackage{soul, xcolor}
\usepackage{mathtools}
\usepackage{siunitx}
\usepackage{textcomp}
\usepackage{amsmath}
\usepackage{enumitem }
\usepackage{outline}
\usepackage{subfig}
\usepackage{lipsum}
\hypersetup{
    colorlinks = true,
    urlcolor   = blue,
    citecolor  = black,
}
\definecolor{Rev1}{rgb}{0, 0, 0}

\newcommand{\RomanNumeralCaps}[1]

\title{Coherent organization of passive scalar from a point-source in a turbulent boundary layer}

\author{Isaiah E. Wall, Gokul Pathikonda
  \corresp{\email{gokul.pathikonda@asu.edu}},
  }

\affiliation{School for the Engineering of Matter, Transport, and Energy (SEMTE), Arizona State University}

\begin{document}
\newcommand{\DraftComment}[1]{\texttt{\textcolor{red}{[#1]}}} %
\newcommand{\AuthorAssigned}[2]{\texttt{\textcolor{blue}{[#1 | #2]}}} %
\newcommand{\LittleHeading}[1]{\noindent\textbf{#1:}} %
\newcommand{\Order}[1]{\ensuremath{\mathcal{O}({#1})}} %
\newcommand{\Retau}[0]{\ensuremath{Re_\tau}} %
\newcommand{\RevisionText}[1]{\textcolor{Rev1}{#1}} 
\maketitle

\begin{abstract}
The spatial organization of a passive scalar plume originating from a point source in a \RevisionText{turbulent} boundary layer is studied to understand its meandering characteristics. We focus shortly downstream of the isokinetic injection ($1.5\le x/\delta \le 3$, $\delta$ being boundary layer thickness) where the scalar concentration is highly intermittent, the plume rapidly \textit{meanders}, and \textit{breaks-up} into concentrated scalar pockets due to the action of turbulent structures. Two injection locations were considered: the center of logarithmic-region and the wake-region of the boundary layer. Simultaneous quantitative \RevisionText{acetone} planar laser-induced fluorescence (\RevisionText{Ac-}PLIF) and particle-image velocimetry (PIV) were performed in a wind-tunnel, to measure scalar mixture fraction and velocity fields. Single- and multi-point statistics were compared to established works to validate the diagnostic novelties. Additionally, the spatial characteristics of plume intermittency were quantified using `blob' size, shape, orientation and mean concentration. It was observed that straining, break up and spatial reorganization were the primary plume-evolution modes in this region, with little small-scale homogenization. Further, the dominant role of coherent vortex motions in plume meandering and break-up was evident. Their action is found to be the primary mechanism by which the injected scalar is transported away from the wall in high concentrations (`large meander events (LMEs)'). Strong spatial correlation was observed in both instantaneous and conditional fields between the high concentration regions and individual vortex heads. This coherent transport was weaker for wake-injection, where the plume only interacts with outer vortex motions. A coherent-structure based mechanism is suggested to explain these transport mechanisms.

\end{abstract}

\begin{keywords}
Authors should not enter keywords on the manuscript, as these must be chosen by the author during the online submission process and will then be added during the typesetting process (see \href{https://www.cambridge.org/core/journals/journal-of-fluid-mechanics/information/list-of-keywords}{Keyword PDF} for the full list).  Other classifications will be added at the same time.
\end{keywords}

\section{Introduction}
The evolution of a passive scalar plume in a turbulent boundary layer (TBL) is a problem of significant interest with direct implications to modeling and predicting pollutant transport, estimation of toxicity and flammability of accidental/intentional chemical releases, modeling odor nuisance, modeling atmospheric chemistries, and source triangulation from discrete scalar measurements, etc \citep{CassianiBertagniEtAl-2020}.
For all of these applications, knowledge about the evolution of the concentration statistics (mean and probability density function, PDF, for eg.) and their relation to the turbulence field evolution is critical. For example, the ability to evaluate chronic effects of an anthropogenic scalar on a neighborhood is strongly related to the ability to accurately estimate mean scalar concentration maps. On the other hand, the estimation of acute effects or flammability/explosivity limits of accidental releases requires a more detailed knowledge of the higher order statistics (variance, level-crossing probability, etc.) or the concentration PDF. The evolution of all of these quantities in real world are almost always coupled with atmospheric weather patterns and natural/urban topographies in a highly non-linear and unsteady manner. 

In the current study, we consider the canonical configuration where we focus on the evolution of a passive scalar injected isokinetically from a point source into a moderate Reynolds number ($Re_\tau$) turbulent boundary layer, that sustains prominent coherent structures \citep{Robinson-1991, AdrianMeinhartEtAl-2000,Adrian-2007}. 
We suspect that capturing the interactions between the coherent structures and scalar plume is essential for developing a comprehensive phenomenological model that can account for rare events that deviate significantly from the mean (such as large meander, high concentration, etc.). Further, we limite our discussion and comparisons to laboratory experiments, and the reader is referred to \citet{FernandoZajicEtAl-2010} for an overview of field measurements made in real atmospheric flows. 

\subsection{Overview of point-source scalar transport in boundary layers}
\begin{figure}
    \centering
    \includegraphics[width=1.0\textwidth]{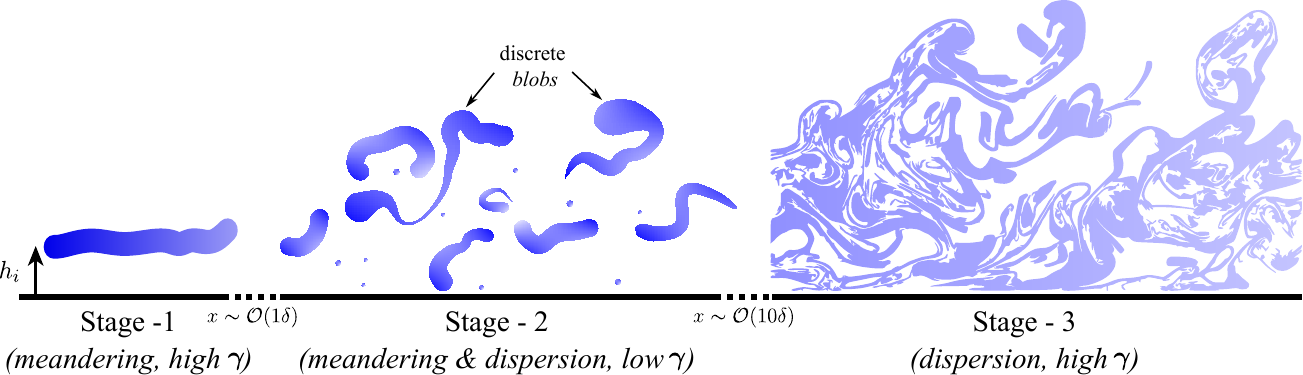}
    \caption{Schematic of different stages of plume evolution and dominant mechanisms (Stage-3 representation based on data from \cite{CrimaldiKoseff-2001}).}
    \label{fig:BlobSchematicV2}
\end{figure}

We can qualitatively describe the evolution of scalar plume injected in a TBL (outlined in figure~\ref{fig:BlobSchematicV2}) from a source with a size $d_s$. In the region immediately downstream ($x\lesssim 1\delta$, \textit{stage 1}, $\delta$ being the boundary layer thickness), the plume is characterized by clear continuous boundaries (demarking the scalar and the ambient fluid) and a homogeneous concentration, that meanders due action of turbulent scales, $\lambda \gg d_s$. Subsequently, the plume stretches, strains and breaks-up in the \textit{intermediate region} ($1\delta \lesssim x\lesssim10\delta$, \textit{stage 2}) that is characterized by discrete parcels of scalar mixing in the TBL. As these parcels grow in wall-normal extent faster than the boundary layer growth, they eventually create a continuous scalar field within the boundary layer (i.e. the boundary layer has approximately \textit{non-zero} concentrations of marked scalar everywhere, \textit{stage 3}). 
This entire process occurs under the influence of a range of turbulent eddies/vortices, and their effects on the plume depends on the relative size of the plume and the eddies within each stage.
In other words, structures \textit{much larger} than the plume predominantly cause the plume to meander, structures \textit{much smaller} than the plume causes mixing within the plume/parcels, and structures of the same order of the plume width tend to stretch and strain the plume (i.e. change the shape of the plume/parcels). The defining quantitative measure of the three evolution stages is the concentration intermittency factor, $\gamma$. At a given spatial region, it is defined as the fraction of time where a non-zero concentration of scalar is observed ($N_{c>0}/N$, where $N$ is an instance of measurement). 
For example, along the injection line in stream-wise direction, $\gamma$ starts at 100\% (where a scalar is always observed, stage-1), decreases to a minimum ({$\approx 40\%$} in region of highly intermittent flow, stage-2), and then increases back to $\approx100\%$ as the scalar is dispersed throughout the boundary layer \citep[stage-3]{FackrellRobins-1982, FackrellRobins-1982a,SawfordFrostEtAl-1985} (note that due to continuous entrainment of free-stream fluid, $\gamma<100\%$ strictly). Thus, high intermittency (low $\gamma$) is a defining characteristic of the intermediate region ($x\approx 1\delta-10\delta$).  

\par
The qualitative nature of the plume-spread in the different stages is a strong function of injection parameters, specifically source size, $d_s$ and the source injection height, $h_i$ within the boundary layer \citep{FackrellRobins-1982, FackrellRobins-1982a, SawfordHunt-1986}.
For a scalar injection very close to the wall ($h_i \approx 0$, known as `ground-level source'), the viscous sub-layer (VSL) has a propensity to collect the scalar over an extended region and serve as a scalar reservoir to the rest of the boundary layer \citep{CrimaldiWileyEtAl-2002}. In contrast, injection at higher elevations tends to spread through the entirety of the boundary layer at a much faster rate, due to the transverse plume growth rate being greater than the boundary layer growth rate \citep{PorehCermak-1964}.
Quantitatively, the evolution of mean scalar concentration from an elevated source is known to follow a self-similar, reflected Gaussian profile (discussed later in equation~\ref{eq:MeanConc}) \citep{FackrellRobins-1982a, YeeKosteniukEtAl-1993,YeeWilsonEtAl-1993, CrimaldiKoseff-2001,CrimaldiWileyEtAl-2002,TalluruHernandez-SilvaEtAl-2017,NironiSalizzoniEtAl-2015}. However, deviations from this form are observed in the outer regions of the turbulent boundary layer (that has both turbulent and non-turbulent regions, \cite{TalluruHernandez-SilvaEtAl-2017}) and very close to the wall \citep{YeeKosteniukEtAl-1993,YeeWilsonEtAl-1993}. Similar Gaussian form was shown to fit the evolution of the concentration variance within the boundary layer as well \citep{CrimaldiKoseff-2001, CrimaldiWileyEtAl-2002}. Self-similarity of temporal scalar spectra was also observed in the spanwise--wall-normal plane when scaled by the outer scales (boundary layer thickness, $\delta$ and free stream velocity, $U_\infty$, \cite{TalluruPhilipEtAl-2019}). 

While the trends for mean and variance of concentration describe the structure of average plume, the turbulent scalar transport mechanisms that result in said trends are captured by the turbulent fluxes ($\overline{u'c'}, \overline{v'c'}, \overline{w'c'}$), measuring which requires simultaneous measurements of both concentration and velocity (here $\cdot'$ quantities represent fluctuations in velocity $u,v,w$ and scalar concentration $c$). The stream-wise and wall-normal turbulent fluxes ($\overline{u'c'}$ and $\overline{v'c'}$ respectively) display a characteristic \texttt{`S'}-shape in their wall-normal variation with an inflection point at the injection height ($y=h_{inj}$). $\overline{v'c'}$ is positive above the injection height and negative below \citep{FackrellRobins-1982a}, consistent with the scalar transport away from $h_i$. Similar trends are also observed for span-wise turbulent flux, $\overline{w'c'}$ in the span-wise direction at all heights.
The steam-wise turbulent flux $\overline{u'c'}$ shows the opposite sign to $\overline{v'c'}$ with an inflection point slightly below $h_i$ for injection locations close to the wall \citep{TalluruPhilipEtAl-2018}. This quantity, however, is not dynamically dominant, relative to cross-stream fluxes (i.e. $\overline{v'c'}$ and $\overline{w'c'}$) since advection due to mean velocity is significantly stronger than turbulent transport \citep{FackrellRobins-1982a}.
Finally, in addition to these trends in single-point statistics, the spatial structure of the plume was quantified via two-point scalar correlation using planar measurements by \citet{Miller-2005} and synchronized two-probe measurements by \citet{TalluruChauhan-2020}, who showed that 2-D two-point scalar correlation map had an oval shape that tilts in the direction of the mean shear. The inclination angle was also found to increase with downstream distance \citep{Miller-2005}. 

\par
These early canonical plume transport studies were primarily directed at developing and validating novel analytical approaches and numerical schemes applicable to realistic scalar transport processes \citep{CassianiBertagniEtAl-2020}. Specifically, the ability to predict spatial evolution of the mean concentration and higher-order statistics is of interest.
One of the earliest mechanistic models that described the evolution of a scalar plume was the `meandering plume' model, first proposed by \cite{Gifford-1959}. The model considers that the plume evolution can be phenomenologically described as the sum of two statistically independent processes: (1)~the meandering of the plume caused by velocity structures asymptotically larger than the plume's width, and (2)~dispersion and mixing at the boundaries of the plume caused by turbulence structures smaller than the width of the plume. 
The model has limitations, namely it starts to break down close to the wall (due to strong mean shear invalidating the turbulence isotropy assumption), far from the source (due to unrestricted growth of the plume, \cite{FackrellRobins-1982a}), and outside the fully turbulent region of the boundary layer $>\sim0.6\delta$ \citep{MarroNironiEtAl-2015,TalluruPhilipEtAl-2018}.
However, in the intermediate regime (stage-2) the plume has been observed to follow the presumed meandering behavior very well, with large eddies making the plume meander, small eddies causing mixing at the boundaries, and eddies roughly the size of the plume causing it to break-up, stretch and elongate \citep{YeeKosteniukEtAl-1993,YeeGailisEtAl-2003, TalluruPhilipEtAl-2018}. For this reason, variations to this approach, collectively known as `fluctuating plume models', have been developed extensively \citep{CassianiBertagniEtAl-2020}.
The meandering plume description was further quantified by \citet{TalluruPhilipEtAl-2018} in explaining the variations in observed turbulent fluxes, concentration time scales, and organization of structures relative to the plume centerline. 
For regions far from the injection location, the action of turbulence on plume mixing is more amenable to analytical approaches such as empirical and semi-empirical modeling, Lagrangian stochastic modeling, etc. Additionally, numerical methods such as RANS and PDF models, and scale resolving simulations such as large eddy simulations (LES) and direct numerical simulations (DNS) are also feasible, but at an increased computational effort. \cite{CassianiBertagniEtAl-2020}, \cite{MarroNironiEtAl-2015} and \cite{MarroSalizzoniEtAl-2018} summarize the state of the art of these numerical efforts. Since most of these models focus on estimating the statistical representations of the concentration fields (mostly pdf and higher order moments of concentration) without explicity considering the turbulent mechanics, we do not discuss these approaches in further detail. The numerical efforts investigating the turbulent structure of the transport processes are discussed in section~\ref{sec:intro-coherent-structures}.

\par
Experimentally understanding the two- and three-dimensional structure of both scalar and velocity, and their interplay, requires multi-point measurements/correlations, either via arrays of probes, via flow-field measurements, or scale resolving numerical simulations. This is particularly important to understand the precise turbulent mechanisms that result in this transport.
Planar Laser Induced Fluorescence (PLIF) is commonly implemented to capture scalar concentration fields in a plane, either in stream-wise--wall-normal ($x-y$) plane \citep{CrimaldiKoseff-2001,CrimaldiWileyEtAl-2002,LiaoCowen-2002}, span-wise--wall-normal ($y-z$) plane \citep{CrimaldiWileyEtAl-2002,VanderwelTavoularis-2016} or stream-wise--span-wise ($x-z$) planes \citep{CrimaldiWileyEtAl-2002,ConnorMcHughEtAl-2018,EismaWesterweelEtAl-2021}.
These studies confirmed the self-similar behavior of concentration and fluctuation profiles in both water \citep{CrimaldiWileyEtAl-2002} and in air \citep{ConnorMcHughEtAl-2018}. 
Lastly, \citet{LiaoCowen-2002} used PIV and LIF to characterize the turbulent structures responsible for the creation and evolution of filament structures in the intermediate downstream region observing high levels of intermittency far from the injection location for scalars with low molecular diffusivity.

\par
\subsection{Coherent structures and scalar transport} \label{sec:intro-coherent-structures}
Beyond the single-point measures, understanding the role of coherent structures that populate the boundary layer flow \citep{Robinson-1991} in the scalar transport requires capturing the scalar and turbulent velocity flow fields simultaneously. 
These coherent structures have a dynamically significant footprint \citep{Adrian-2007}, and a finite spatio-temporal coherence (i.e. they exist for a long enough time in the flow to impact the flow qualitatively).
Prominent structures of interest are the quasi-streamwise vortices close to the wall, wall-attached hairpin vortices in the logarithmic region \citep{KlineReynoldsEtAl-1967, Townsend-1976}, and hairpin packets that form a coherent group of hairpin vortices \citep{AdrianMeinhartEtAl-2000, Adrian-2007}. 
The role of near-wall quasi-streamwise vortices on scalar transport has not been extensively studied since they are dominant only over hydrodynamically smooth surfaces, with limited relevance to atmospheric flows (that are typically `fully rough'). 
On the other hand, hairpin vortices and hairpin packets populate the outer regions of a high-$Re_\tau$ ($\gtrsim 1000$) boundary layer over both smooth- and rough-surfaces \citep{VolinoSchultzEtAl-2007, WuChristensen-2007}. The term `hairpin vortex' here will encompass all similar structured (hairpin, horseshoe, omega) vortices and their asymmetric and/or incomplete counterparts. 
They were first theorized by \citet{Theodorsen-1952}, and their dynamical significance has been increasingly evident in recent times in creating large regions of roughly uniform momentum, generation of turbulent kinetic energy in the outer regions, transporting the near-wall turbulence via intense bursting events, etc. \citep{Townsend-1976,Robinson-1991,AdrianMeinhartEtAl-2000,ChristensenAdrian-2001, GanapathisubramaniLongmireEtAl-2003, GanapathisubramaniHutchinsEtAl-2005, NatrajanChristensen-2006,Adrian-2007}. Further, these structures are dynamically dominant only at high-$Re_\tau$ conditions (i.e. $Re_\tau \gtrsim 1000$). 

\par
Efforts to study the role of these large structures on the scalar transport are relatively recent. \citet{VanderwelTavoularis-2016} conducted stereo-PIV and PLIF in a plane-shear flow (without solid boundaries) and, using conditional averages, identified the structure of scalar transport by individual hairpin-type vortices.
They found that both upright and inverted hairpin vortices preferentially oriented the scalar at angles of $155\degree$ and $-25\degree$ with respect to the flow direction, which correlates with an ejection/sweep events of the respectively aligned hairpin vortices (unlike plane shear flows, TBLs have almost exclusively upright hairpin vortices). 
This is similar to the sub-filter scale turbulent energy transfer and scattering events observed in boundary layers \citep{YeeKosteniukEtAl-1993, NatrajanChristensen-2006}. 
Following this, \citet{TalluruHernandez-SilvaEtAl-2017, TalluruPhilipEtAl-2018} performed two-probe simultaneous measurements within a high-$Re_\tau$ boundary layer and found preferential organization of the coherent structures relative to the plume structure and the presence of low- and high- speed structures above and below the plume respectively. The researchers expanded on these observations in their subsequent works \citep{TalluruPhilipEtAl-2019,TalluruChauhan-2020} to investigate the structural inclination angle, self-similarity in temporal spectra and noticed a strong outer-layer scaling. They schematically provided a representative arrangement between momentum and scalar plume. Finally, a recent investigation by \cite{EismaWesterweelEtAl-2021} using high-speed, simultaneous measurements of scalar plume (in a plane using PLIF) and turbulent velocity field (in a volume, using volumetric PIV) identified existence of uniform concentration zones (UCZs) similar to that of uniform momentum zones (UMZs) sustained by the hairpin vortex packets \citep{Adrian-2007}. Additionally, the Lagrangian analysis using Finite Time Lyapunov exponents (FTLE) identified a strong correlation with the UCZ boundaries.
These measurements correspond to downstream sections away from the injection location where the plume already occupies a significant portion of the boundary layer (i.e. Stage-3 in figure~\ref{fig:BlobSchematicV2}). Finally, even though not directly investigating a passive scalar from a point source, \citet{SaleskyAnderson-2020, DharmarathneTutkunEtAl-2016} and \citet{LaskariSaxton-FoxEtAl-2020} all identified the role of coherent structures in wall-normal turbulent transport processes, and \cite{SaleskyAnderson-2020} proposed a model to capture their modulating influence in the flux-gradient relationship.

\subsection{Current work}
The current work focuses on building on this existing understanding and further elucidating the role of the turbulent structures on influencing the local transport of a passive scalar. The previous experimental studies that have explicitly looked into the role of coherent structures and high-$Re_\tau$ effects were either point measurements \citep{TalluruHernandez-SilvaEtAl-2017, TalluruPhilipEtAl-2018,TalluruPhilipEtAl-2019,TalluruChauhan-2020} or flow field measurements performed in water (due to diagnostic simplicity, \cite{VanderwelTavoularis-2014, VanderwelTavoularis-2016,EismaWesterweelEtAl-2021}). The former provide limited spatial information, while the latter have high Schmidt number, $Sc$ effects at high wavenumbers \citep{BuchDahm-1996, BuchDahm-1998}. Additionally, the previous measurements predominantly focused on plume structure and mixing in Stage-3 where the plume occupies a significant fraction of the boundary layer. We hypothesize that the action of coherent structures in the region immediately downstream of the injection point (stage-2, where the injected plume destabilizes into discrete scalar pockets) will have a significant impact and footprint on the downstream evolution into a continuous scalar field. To this end, the current work aims to provide a spatial description of the discrete plume structure, intermittency, and arrangement between $1\delta-3\delta$ from the injection point. To facilitate this, we implement quantitative acetone-PLIF synchronously with PIV in a high-Reynolds number boundary layer. This technique is similar to that in \cite{ConnorMcHughEtAl-2018}, but for controlled boundary layer injection within the boundary layer, and measured simultaneously with the velocity flow-field information.

\section{Experimental Setup}
\label{sec:Expsetup}
\subsection{Facility and Diagnostics}
\begin{figure}
    \includegraphics[width=1.0\textwidth]{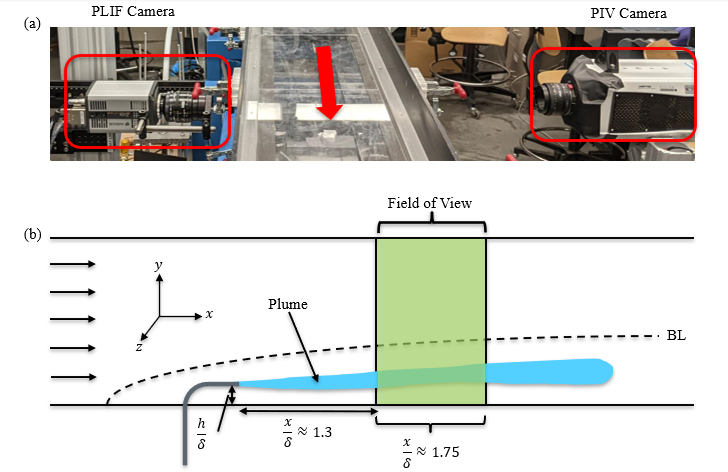}
    \caption{(a) Top view of wind tunnel setup and cameras, (b) schematic side view of experimental setup. The field of view captures stage-2 evolution from figure~\ref{fig:BlobSchematicV2}.}
    \label{fig:ExperimentalSetup}
\end{figure}
The experiments were conducted in a low speed wind tunnel with a cross section of \(152\,\mathrm{mm} \times 152\,\mathrm{mm}\) and a length of 1.42\,m. The boundary layer grows on the smooth, bottom surface of the wind-tunnel, and is tripped using a cylindrical rod. A small tube with a diameter of \(d_i/\delta=0.026\,(0.625\,mm)\) was used to inject an acetone-air mixture at two heights $h_i/\delta=0.135$ and $h_i/\delta=0.307$ which corresponds to the boundary layer log- and wake- regions respectively. We refer to these two injection cases as \textit{log-injection} and \textit{wake-injection} cases in this paper. This mixture was injected isokinetically \RevisionText{at a velocity of $u_{inj}$ at the two injection points such that $u_{inj}=\overline{U}(y_{inj})$.} Measurements were made starting $1.3\delta$ downstream from the injection point, with the field of view spanning  \(1.4\delta \times 2.0\delta\) in stream-wise--wall-normal plane. The schematic shown in figure~\ref{fig:ExperimentalSetup} illustrates the general experimental setup.

\par
To measure the flow field, simultaneous measurements of scalar and velocity fields were performed using 2-component PIV and Acetone PLIF, with synchronized imaging performed from the opposite sides of the laser sheet (figure~\ref{fig:ExperimentalSetup}a).
A single dual-cavity Quantel Q-Smart Nd:YAG laser provided both 532\,nm ($380\,mJ$/pulse) and 266\,nm ($90\,mJ$/pulse) laser pulses at 10Hz required for the simultaneous measurements. An extensive optical setup was necessary to separate and re-combine the desired wavelengths from the Nd:YAG laser for optimal performance. Since the raw quadrupled-Nd:YAG beam contains \(1064\,\mathrm{nm} + 532\,\mathrm{nm}+266\,\mathrm{nm}\), a series of optics were used to isolate the \(532\,\mathrm{nm}+266\,\mathrm{nm}\) wavelengths and dump the \(1064\,\mathrm{nm}\) into a beam block. The combined \(532\,\mathrm{nm}+266\,\mathrm{nm}\) beams were then focused through a series of spherical lens and a sheet-forming cylindrical lens. This focused the laser into a thin sheet in the stream-wise--wall-normal plane and was projected into the wind tunnel from the optical setup underneath. Lastly, a BNC Model 575 timing box was used to control the timing of the flash-lamp and Q-switch delay for both laser heads, and the exposure window for both cameras. The PIV camera captured images from both laser pulses while the Ac-PLIF camera captured only the first pulse. 
\par
The mie-scattered light was used to perform planar PIV in the field of view shown in figure \ref{fig:ExperimentalSetup} using a 4\,MP Phantom V641 camera imaging through a $532\,$nm band-pass filter. A laskin nozzle filled with olive oil provided the seeding particles necessary for the PIV. The PIV image processing was performed using LaVision Davis 10 with a recursive, multi-pass refinement to a final interrogation window size of $16 \times 16\,px^2$ with 50\% overlap. Lastly, in post-processing, outlier vectors in the PIV field were filled in using interpolation from the surrounding vectors.
\par
For the injected scalar, acetone-air mixture was created using an acetone bubbler apparatus consisting of a vertical steel pipe submersed in an acetone bath. Pressurized air entering from the top end of the pipe is forced through small holes at the bottom \RevisionText{end of the pipe} submerged \RevisionText{in} an acetone bath. A heater is wrapped around the acetone bath to increase the temperature and the acetone saturation concentration of the injected plume. The resulting acetone/air mixture created from the bubbler was diverted into two streams. One of the streams was controlled electronically using a MKS-GM50A mass flow controller, and injected into the boundary layer through the injection tube. The excess acetone was vented away. This way, the injected acetone/air mixture is controlled such that the injected plume is isokinetic with the incident boundary layer flow.
The Ac-PLIF was performed by exciting the injected acetone gas with the \(266\,\mathrm{nm}\) laser sheet. The fluoresced light from injected acetone was then imaged with an Andor Zyla sCMOS camera with a 500 nm short-pass filter. The use of this filter greatly reduced the stray fluorescence and reflections from the 266\,nm and 532\,nm light by the surrounding materials (mainly the walls/structure of the wind tunnel). The acetone concentration was small enough that absorption of the laser energy across the scalar field is minimal, and we could use an `optically-thin' assumption \citep{Crimaldi-2008}. This avoids the need for corrections of Beer-Lambert attenuation.
The field of view of both the PIV and PLIF camera overlap and both fully capture the laser sheet. To spatially calibrate both fields and overlay them on the same physical space, a transparent calibration target visible to both cameras was imaged in the location of the laser sheet. This allowed the PIV and Ac-PLIF data to be projected onto the same plane and simultaneous fields to be calculated. The final calibration and magnification resulted in a vector field resolution of $0.53\,mm$ and a scalar field resolution of $0.078\,mm$.
\subsection{Corrections to Acetone PLIF}
\label{sec:PLIF_corrections}
The raw fluorescence images, $I_e$ (around 2640 uncorrelated, instantaneous snapshots) require several corrections to convert image intensities to quantitative mixture fraction, $\xi$ measurements. 
In addition to the fluorescence images of the injected plume, background (BG) and laser-sheet profile, $I_{cal}$ images were also taken after each experiment. Background images were taken in the same conditions as that of the actual experiment but with no acetone being injected. This accounts for laser scatter and stray fluorescence/reflections. The laser sheet profile images were taken by creating a stagnant chamber of acetone vapor of uniform concentration in the wind tunnel, and taking a small ensemble of fluorescence images.
\par
\begin{figure}
    \centering
    \includegraphics[width=1.0\textwidth]{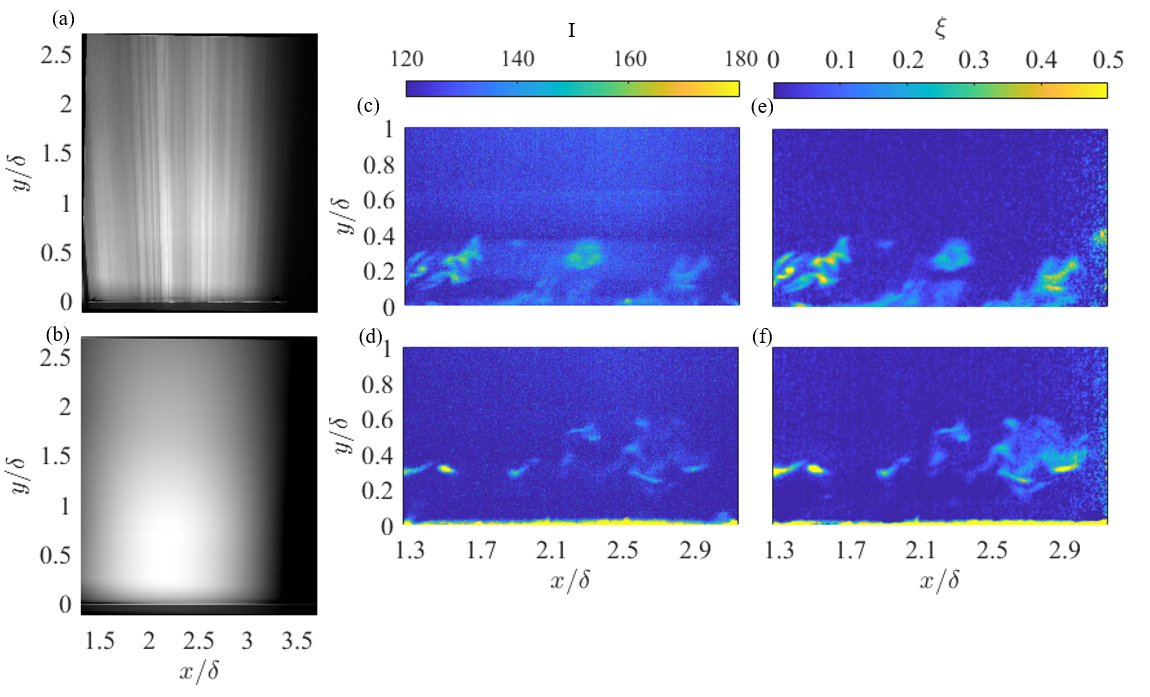}
    \caption{(a)~Calibration laser intensity image, $I_{cal}$ (b)~\textit{ideal} laser intensity field, LS (c)~and (e)~show example fluorescence image (I) and mixture fraction field ($\xi$) respectively for the log-injection case. (d)~and (f)~show the same for wake-injection case. }
    \label{fig:PLIFpics}
\end{figure}

 
It must be noted that due to the large laser energies involved and the presence of two wavelengths (380\,mJ at 532\,nm + 90\,mJ at 266\,nm), the anti-reflective coating and the optics, especially the sheet-forming cylindrical lens, were found to be susceptible to minor damage during the experiment. Due to this, the laser sheet develops striations and local changes in energy as is shown in figure~\ref{fig:PLIFpics}a that translate into noise on turbulent statistics. It was found that this can be compensated to an extent by applying a row-wise parabolic fit to the laser sheet profile that provides an `ideal sheet' (LS) (figure~\ref{fig:PLIFpics}b) and accounts for these local imperfections. This was used in lieu of $I_{cal}$ to correct for spatial variations in laser energy.

Finally, after excluding a handful of outlier images (due to laser pulse malfunction), each of the fluorescence images (I) were corrected for background (BG) and laser energy (LS) variations to compute the acetone concentration fields (relative to acetone mole fraction in a calibration experiment), and scaled to injection concentration ($I_0$) to yield the mixture fraction fields, $\xi$ a shown in equation~\ref{eq:PLIFcorr}.
\begin{equation}
    \xi=\frac{I-BG}{LS-BG}*\frac{1}{I_0}
    \label{eq:PLIFcorr}
\end{equation}
Examples of the LS field, two illustrative initial fields, I, and the corresponding mixture fraction fields, $\xi$ are shown in figure \ref{fig:PLIFpics} for log- and wake- injection cases.
Median and Gaussian filtering (3 pixels) were used to remove the speckle and high frequency noise respectively. The final fields were mapped on to world coordinates using LaVision Davis 10 based on the calibration image so that they overlay on to the PIV vector field down to pixel resolution.
\par

\subsection{Boundary Layer and Injection Parameters}
\begin{wraptable}{r}{0.55\textwidth}
    \caption{Incident boundary layer characteristics} 
    \begin{tabular}{ccc}
    \label{tab:tbl-parameters}
    \textbf{Description} & \textbf{Parameter} & \textbf{Value}                        \\
    Injection height &$h_i/\delta$         & \begin{tabular}[c]{@{}c@{}}0.135 (Log-), \\ 0.307 (Wake-)\end{tabular} \\
    Freestream velocity & $U_\infty$         & $10.67\,m/s$                                   \\
    Skin-friction $Re$ & $Re_\tau$          & 972                                               \\
    BL thickness & $\delta_{99}$           & $24\,mm$                             \\
    Friction velocity & $u_\tau$           & $0.48\,m/s$                          \\
    Inner length scale & $y^*$              & $3.52 \times 10^{-5}$                           \\
    Injector diameter & $d_i/\delta$       & 0.026                                         
    \end{tabular}
\end{wraptable}
Preliminary PIV data \RevisionText{were} gathered for 3 different cases to characterize the incident boundary layer, ensure isokinetic injection of the acetone, capture the effects of the physical presence of the injector on the boundary layer, and the presence of acetone on the boundary layer profile. This was to ensure that the injector pipe and acetone injection would not significantly alter the boundary layer physics. This was conducted for both the log-injection, and the wake-injection in three PIV experiments performed in the FOV of interest:  
(1)~first, the unperturbed boundary layer without the injector was measured. This will be referred to as the `baseline case'. Then, (2)~the injector was positioned at respective heights without injecting any acetone and the boundary layer profiles were measured. This will be referred to as the `no injection' case. Finally, (3)~injector was positioned at respective heights and acetone was injected under isokinetic conditions. This will be referred to as the `isokinetic injection' case, which is the focus of the the current study. 
\par

\begin{figure} 
    \centering
    \includegraphics[width=1.0\textwidth]{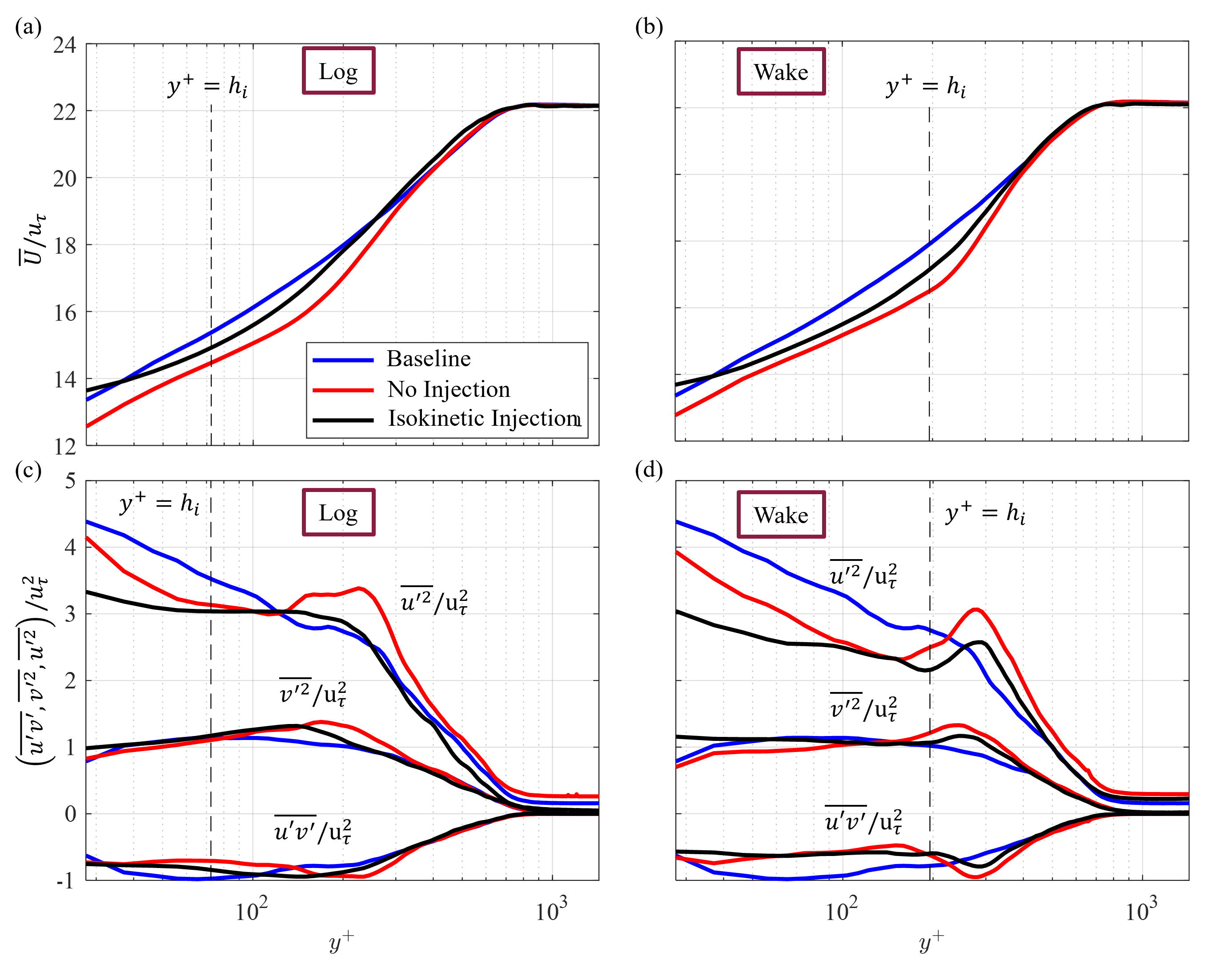}
    \caption{Boundary layer profiles (a,b) and Reynold's stresses (c,d) for both injection locations}
    \label{fig:BLProfiles}
\end{figure}

The mean boundary layer profiles are shown for all three cases for the log and wake injection in figure \ref{fig:BLProfiles}(a,b). 
For both injection locations, the `no injection' case seems to shift the velocity profile downwards from the wall until around $y^+=300$ due to blockage of the injector stem. The `isokinetic injection' case seems to help the boundary layer recover and more closely resemble the canonical boundary layer compared to the `no injection' case. 
Except for minor differences, the boundary layer profiles follow the same trend for baseline and the isokinetic injection cases.
\par
We also compare the changes in in-plane Reynold's stresses, as shown in figure \ref{fig:BLProfiles}(c,d).
The injection, no flow case, and the isokinetic injection case for the stream-wise Reynold's stress deviate from the canonical case close to the wall, possibly due to the wake of the vertical injector stem.
Away from the wall, the values of all three cases converge on the canonical value for the stream-wise stress. Interestingly, there seems to be a jump in all three stresses at a location just before $y^+=200$ for both injection locations. This jump is more prominent and noticeable for the wake injection. Nonetheless, for both cases the stresses quickly converge back to a common value consistent with the canonical profile. Overall, the wake injection has more persistent deviation from the canonical case than the log injection, and for both cases the stream-wise Reynolds stress has the greatest deviation from the canonical profile. Despite these deviations, the overall trends of the three stresses for canonical and isokinetic injection cases are very similar. 
\par
The relevant boundary layer characteristics for the canonical case are summarized in table~\ref{tab:tbl-parameters}, and these values are used for all normalizations in the current work. The boundary layer inner-scale parameters, namely $u_{\tau}$, $y^*$, and $Re_\tau$, were estimated by finding the best-fit of the measured mean boundary layer profile to zero-pressure gradient composite profile proposed by \cite{ChauhanMonkewitzEtAl-2009}. 
\subsection{Simultaneous PIV-PLIF Snapshots}
\begin{figure}
    \centering
    \includegraphics[height=0.95\textheight, trim={0in, 0.5in, 0in, 0in}]{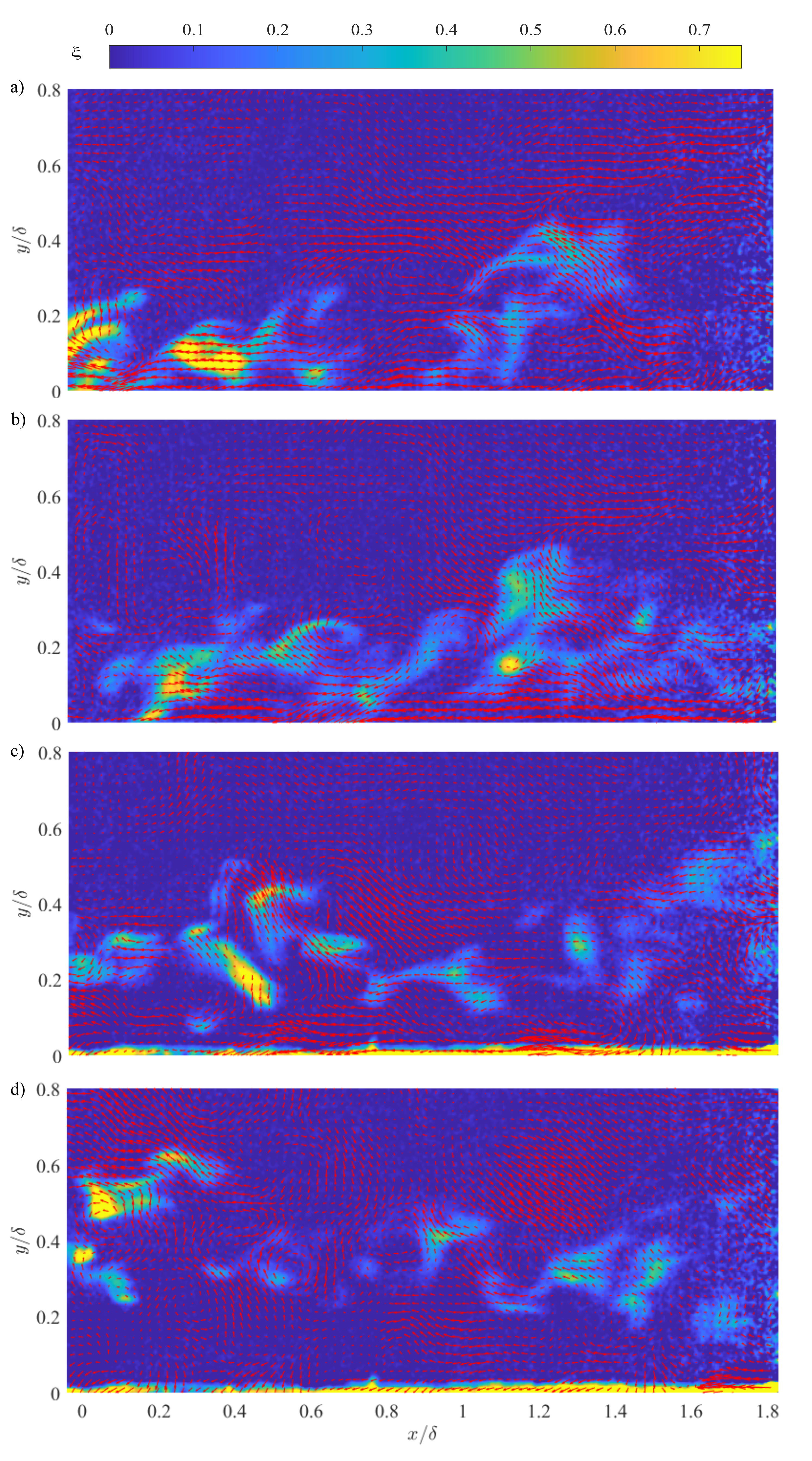}
    \caption{Instantaneous snapshots of simultaneous mixture fraction ($\tilde{\xi}$) and velocity fields for (a,b)~log- and (c,d)~wake-injection cases.}
    \label{fig:InstantSnap1}
\end{figure}

Four representative instantaneous fields are shown in figure~\ref{fig:InstantSnap1} that display the overlapped velocity and mixture fraction data for log- and wake- injection. Gaussian low and high pass filters of lengths $1.25\,\delta$ and $2\times$vector spacing respectively, were applied to the velocity field to filter out the large-scale local advection and high-frequency noise \citep{AdrianChristensenEtAl-2000a}. This demonstrates the ability to perform simultaneous field measurements of quantitative scalar mixture fraction and turbulent velocity flow fields and qualitatively highlights the turbulent dynamics responsible for scalar transport within the boundary layer.  We utilize this capability to capture the spatial organization of the scalar dissipation relative to the coherent motions of the boundary layer akin to those obtained by point measurements by \citet{TalluruPhilipEtAl-2018} and \cite{TalluruChauhan-2020}, and calculate scalar two-point auto- and cross-correlations using simultaneous spatial measurements.

\section{Mean Plume Evolution and Analytical Models}
The mean values of $u,v,$ and $\xi$ in the plume were calculated by taking the ensemble average of the roughly 2600 fields after the corrections have been applied described in section \ref{sec:Expsetup}. These mean values were then used to calculate the perturbation fields of $u,v,$ and $\xi$ and turbulent fluxes in each field. 

\begin{figure}
    \centering
    \includegraphics[width=1.0\textwidth]{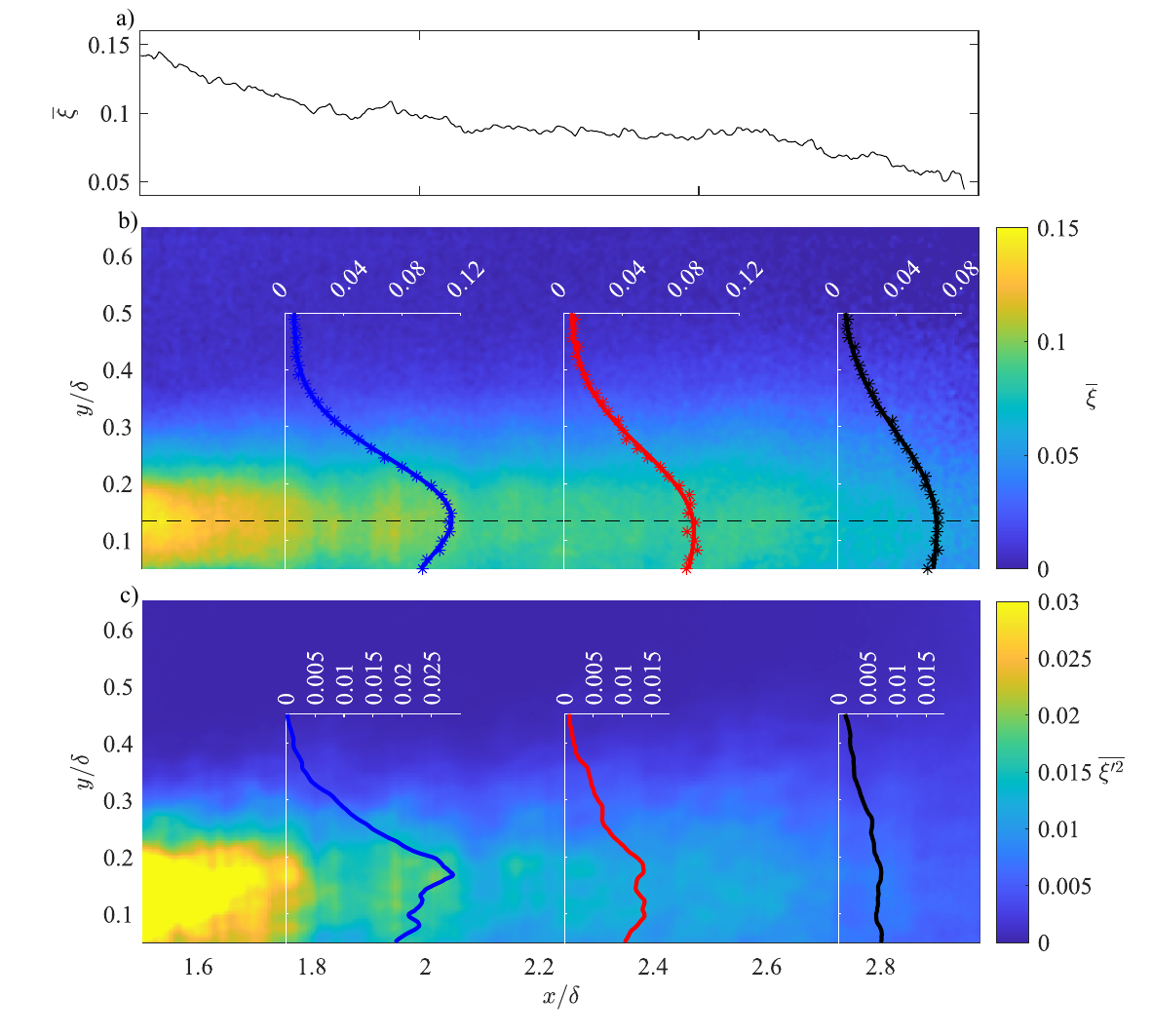}
    \caption{Log-injection case: (a)~mean mixture fraction on injection line $\overline{\xi}\left( {y=h_{i}} \right)$, (b)~mean mixture fraction $\overline{\xi}$ field, and (c)~variance of mixture fraction field $\overline{\xi'^2}$}
    \label{fig:LogMeanVarplts}
\end{figure}

\begin{figure}
    \centering
    \includegraphics[width=1.0\textwidth]{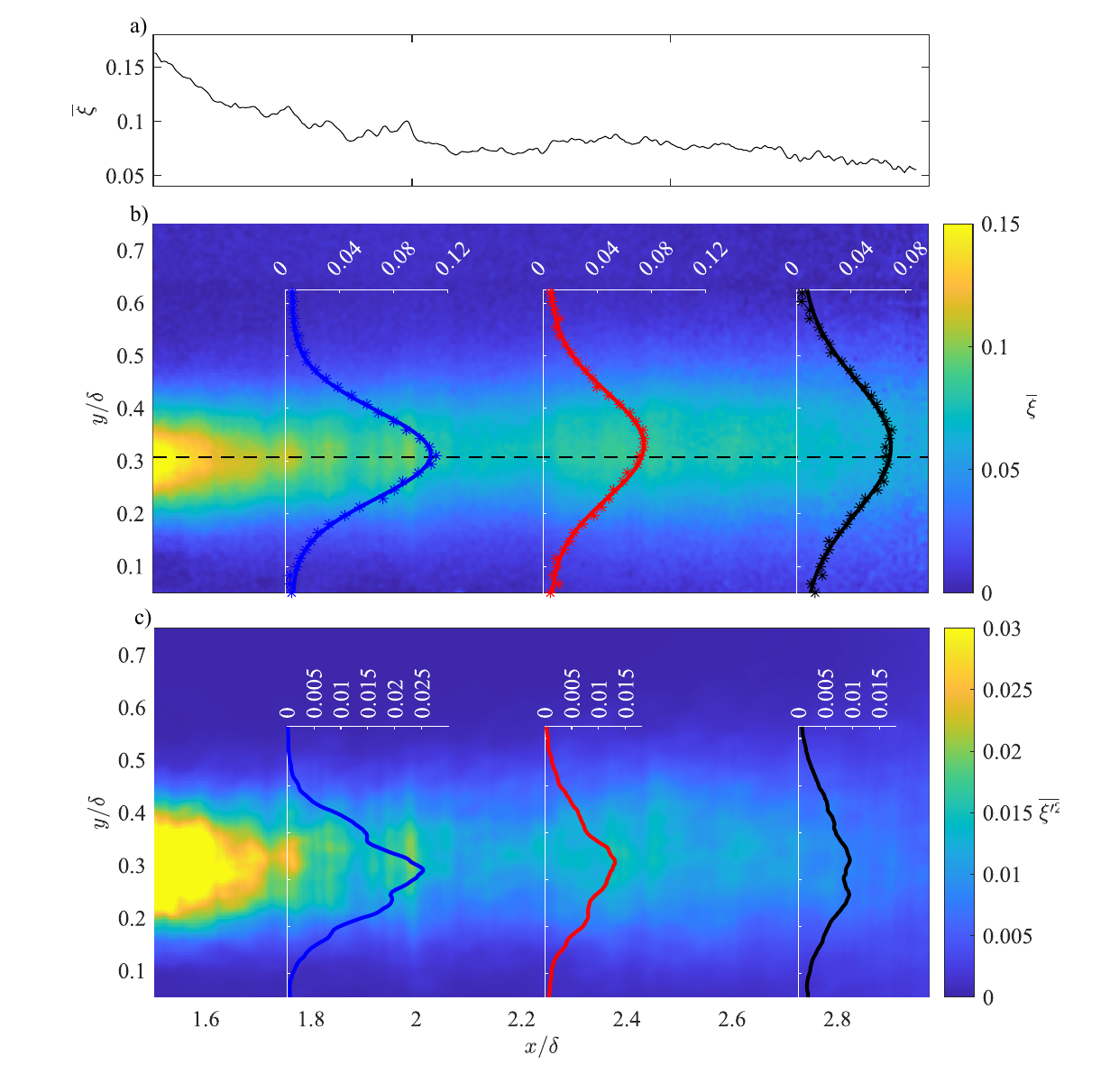}
    \caption{Wake-injection case: (a)~mean mixture fraction on injection line $\overline{\xi}\left( {y=h_{i}} \right)$, (b)~mean mixture fraction $\overline{\xi}$ field, and (c)~variance of mixture fraction field $\overline{\xi'^2}$}
    \label{fig:WakeMeanVarplts}
\end{figure}

The evolution of the mean concentration profile as a function of downstream location follows a reflected Gaussian function of form equation~\ref{eq:MeanConc} first proposed by \cite{FackrellRobins-1982a}.
Figures \ref{fig:LogMeanVarplts} and \ref{fig:WakeMeanVarplts} show the mean and variance plots for the log and wake injection respectively with a line plots of the contour shown at three stream-wise locations. 
For the mean concentration line profiles, equation \ref{eq:MeanConc} is also shown fitted to the mean concentration data at the chosen downstream locations showing an excellent fit for both injection locations. The variance plots for both the log- and wake- injection both seem to follow a rough reflected Gaussian distribution with a peak at the injection height, as was previously observed by \cite{FackrellRobins-1982a} and \cite{Miller-2005}.

\label{sec:PSAM}
Many previous studies have utilized single point measurements to understand plume dispersion in TBLs. \citet{NironiSalizzoniEtAl-2015} compared experimentally measured plume dispersion with analytical models based on hypotheses from \citet{FackrellRobins-1982a, FackrellRobins-1982} with good agreement, the procedure for which will be followed here. 
This equation assumes the interaction of the plume with the wall and accounts for this via a virtual source below the wall at a wall normal location of $y=-h_{i}$, given as
\begin{equation}
    \overline{c}(x,y)=\frac{M_q}{2\pi \sigma_y \sigma_z \overline{u}_{adv}}\left[exp\left(-\frac{(y+h_i)^2}{2\sigma_y^2} \right)+exp\left(-\frac{(y-h_i)^2}{2\sigma_y^2} \right) \right]
    \label{eq:MeanConc}
\end{equation}
With $M_q$ denoting the mass flux from the injector, $\sigma_y$ and $\sigma_z$ denoting the plume spread in the wall-normal and span-wise directions respectively, and $h_s$ denoting the injection height of the plume. By using the best-fits to the mean-profile (figures~\ref{fig:LogMeanVarplts} and~\ref{fig:WakeMeanVarplts}), the various parameters can be extracted. However, since only the wall normal plume spread $\sigma_y$ is available \RevisionText{from the PLIF measurements}, the span-wise plume spread, $\sigma_z$, is approximated as $\sigma_z\approx\sigma_y$. \RevisionText{This assumption is necessary to calculate all the variables in equation \ref{eq:MeanConc}.} This approximation holds reasonably well for elevated sources as was illustrated by \citet[Fig.~6]{NironiSalizzoniEtAl-2015} (though the wall-normal plume is typically slightly less than the spanwise spread).

This wall-normal plume spread, $\sigma_y$, can also be estimated analytically via the Lagrangian transport and dispersion of the scalar plume.
A Lagrangian plume, growing in time, $t$ in a turbulent environment can be described as \RevisionText{(see \cite{FackrellRobins-1982a})}.
\begin{equation}
    \sigma_y^2=\frac{\sigma_0^2}{6}+2\sigma_v^2T_{Lv}\left(t-T_{Lv}\left[1-\exp\left(-\frac{t}{T_{Lv}}\right)\right]\right)
\label{eq:PlumeDelta}
\end{equation}
With $\sigma_0$ being the equivalent initial plume diameter (typically $=d_i$), $\sigma_v^2$ is the variance of the turbulent velocity fluctuation, and $T_{Lv}$ denoting the Lagrangian time scale of the turbulent environment. 
The time in equation \ref{eq:PlumeDelta} can be converted to spatial distance using Taylor's frozen field hypothesis (i.e., $t=x/U$).
Assuming isotropy in turbulence, the Lagrangian time scale $T_{Lv}$ can be approximated by the Eulerian \RevisionText{integral length scale} $L_{vv}$ using equation \ref{eq:LangTS}
\begin{equation}
    T_{Lv}=\frac{L_{vv}}{\sigma_v^2}=\frac{L_{vv}}{\overline{{v'}^2}}
\label{eq:LangTS}
\end{equation}
With $\overline{{v'}^2}$ denoting variance of velocity perturbation in the wall-normal direction. Finally, the integral transverse length scale, $L_{vv}$ can be estimated by fitting an exponential decay to the auto-correlation of $v'$ in the wall-normal direction. More details can be found in \cite{NironiSalizzoniEtAl-2015}. 

\begin{figure} 
    \centering
    \includegraphics[width=0.9\textwidth]{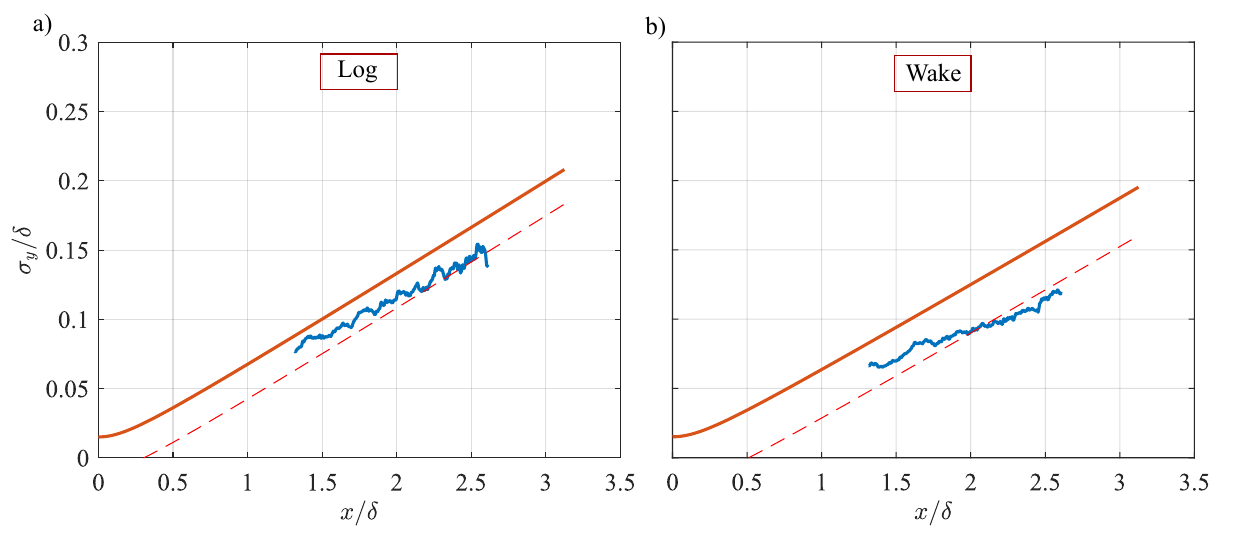}
    \caption{Comparison of theoretical plume spread (\textcolor{red}{red}) predicted by equation \ref{eq:PlumeDelta}, experimental plume spread (\textcolor{blue}{blue}) obtained by fitting equation \ref{eq:MeanConc} to the current data, and theoretical plume-spread offset vertically for reference (\textcolor{red}{- -})}
    \label{fig:PlumeFitExpTheo}
\end{figure}

Since the turbulent velocity fields were measured, the properties of equation \ref{eq:PlumeDelta} can be calculated to form the theoretical plume spread, and compared against the $\sigma_y$ measured experimentally. This comparison is shown in figure~\ref{fig:PlumeFitExpTheo} for both log- and wake-injection cases.
It can be seen that the overall trends of the plume spread is accurately captured by the model, with some discrepancies. The plume in the current experiments appears to be offset from that predicted by the model for the logarithmic region. This can be associated to the weak definition of $\sigma_0$, which is an effective source plume width not always exactly equal to the source diameter, $d_i$. For the wake region, the plume is offset and appears to grower slower than that predicted by the model. The authors suspects this to be from the large size of the injector relative to the boundary layer ($d_i/\delta$) and the limitations of indirectly estimating the Lagrangian time-scales using equation~\ref{eq:LangTS} in the intermittent wake region. It should also be noted that the nondimensionalized injection diameter of the current experiment is an order of magnitude larger than that seen in \citet{NironiSalizzoniEtAl-2015}, which can add finite source-diameter effects not captured by the models.

\section{Turbulent Scalar Fluxes} \label{sec:turbulent-flux}
\begin{figure}
    \centering
    \includegraphics[width=1.0\textwidth]{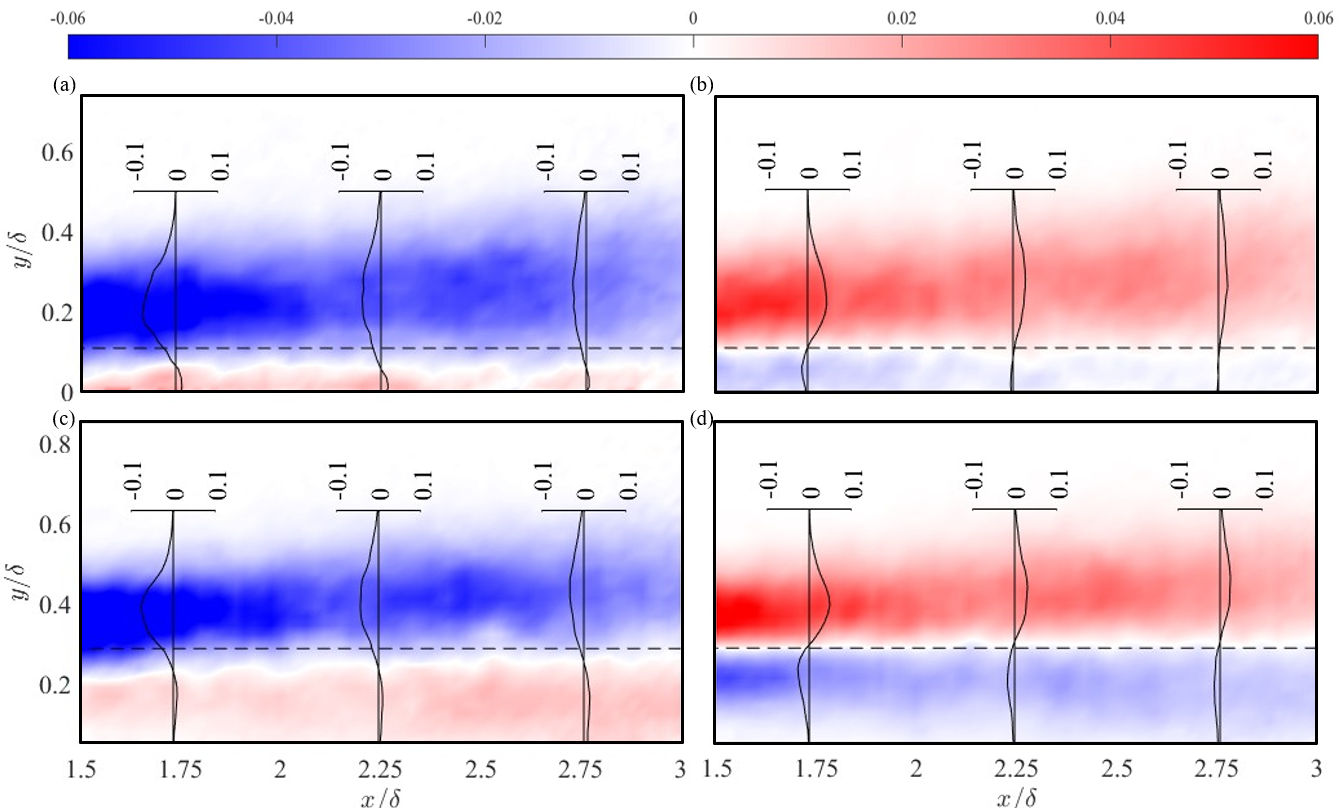}
    \caption{(a,c)~Streamwise ($\overline{u'\xi'}/u_\tau$) and (b,d)~wall-normal ($\overline{v'\xi'}/u_\tau$) turbulence scalar flux for (a,b)~log- and (c,d)~wake- injection cases.}
    \label{fig:TurbFlux}
\end{figure}

The simultaneous measurement of scalar and turbulent velocity fields enables us to compute the turbulent scalar fluxes in the measurement plane (i.e. $\overline{u'\xi'}$ and $\overline{v'\xi'}$). Since the scalar fields are acquired at higher spatial resolution than the velocity fields [$(0.078\,mm)^2$ vs $(0.530\,mm)^2$], the scalar fields are downsampled to the velocity field resolution to estimate these cross-statistics.
Contour and line plots of $\overline{u'\xi'}/u_\tau$ and $\overline{v'\xi'}/u_\tau$ are shown for the log and wake region injection points in figure \ref{fig:TurbFlux}. The contour plots show clear trends in both regions centered around the injection point. For the stream-wise perturbations ($u'$) there are on average positive correlations with $\xi'$ below the injection line and negative correlations above the injection line. The wall-normal flux shows a positive correlation above the injection point and a negative correlation below, mirroring the trend seen for the stream-wise flux. There is also a distinct net zero flux region at the injection line for the wall-normal flux, and just below the injection point for the stream-wise flux. 

These trends are consistent with the existing literature \citep[for eg.]{FackrellRobins-1982,CrimaldiWileyEtAl-2002,NironiSalizzoniEtAl-2015,TalluruPhilipEtAl-2018}, and for a mean plume spreading in wall-normal and spanwise extents. The mean transport of a high concentration pocket of fluid ($\xi'>0$) originally at the injection line away ($v'>0$) and towards ($v'<0$) the wall will have positive and negative values respectively of wall-normal turbulent flux, $\overline{v'\xi'}$. Similarly, the streamwise turbulent flux, $\overline{u'\xi'}$ away from the injector line will have the opposite sign to $\overline{v'\xi'}$ owing to the mean negative Reynolds shear stress in the entire boundary layer, i.e. $\overline{u'v'}<0$.

\subsection{Two-Point Mixture Fraction Correlation}
\begin{figure}
    \centering
    \includegraphics[width=1.0\textwidth]{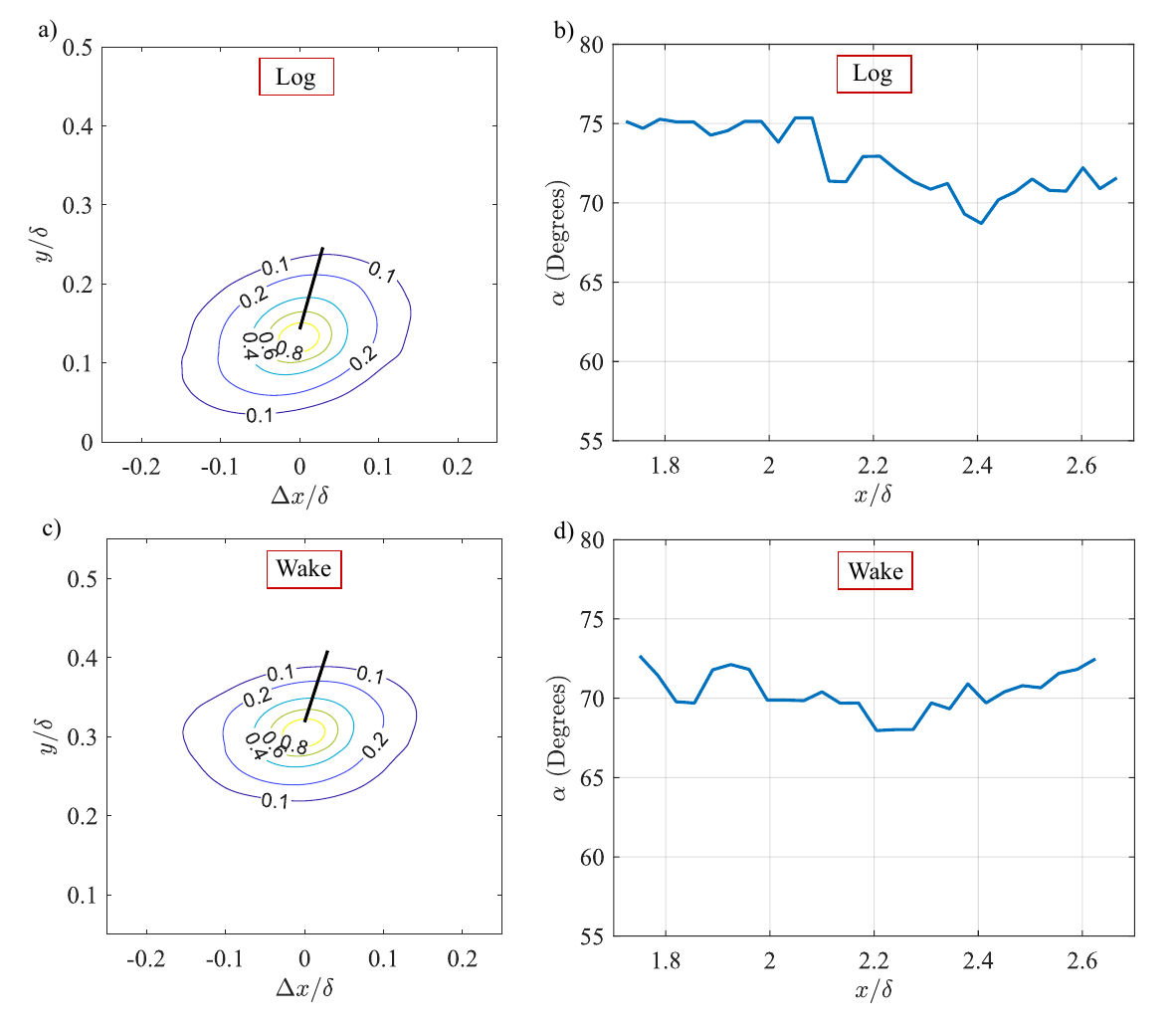}
    \caption{Angle of maximum mixture fraction correlation for (a,b)~log- and (c,d)~wake injection. (a,c)~indicate a illustrative local correlation map at $x/\delta=1.9$ for the two cases.}
    \label{fig:InclinAngle}
\end{figure}

The spatial scalar fields enables us to compute the multi-point turbulent statistics directly, without the need for Taylor's hypothesis.
The two-point mixture fraction correlation, $R_{\xi\xi}$, defined by equation \ref{eq:ccorrelation}, is computed to understand the spatial coherence of the scalar.
\begin{equation}
    R_{\xi\xi}(\Delta x, \Delta_y; x_r,y_r)=\frac{\overline{\xi'(x_r,y_r)\xi'(x_r+\Delta x,y_r+\Delta y)}}{\overline{\left(\xi'(x_r,y_r)\right)^2}}
    \label{eq:ccorrelation}
\end{equation}
These correlations represent the general shape, extent and inclination of the \RevisionText{mixture fraction} neighborhood around $(x_r, y_r)$ (where a certain value of \RevisionText{mixture fraction}, $\xi'$ is detected). Previous studies \citep{Miller-2005, TalluruChauhan-2020} using point-measurements (and Taylor's hypothesis) have shown that the 2-point mixture fraction correlation fields are inclined at an angle with the horizontal, where the location of maximum correlation above the reference location ($y>y_r$) is ahead in streamwise location ($x > x_r$), and vice versa.
To investigate this phenomenon spatially using the current data, the correlation fields were calculated using equation~\ref{eq:ccorrelation} for $y_r = h_i$ at various streamwise locations, $x_r$. Further, to achieve reasonable convergence, these fields were locally averaged over streamwise span of $0.65 \delta$ and $0.7 \delta$ for the log and wake injection respectively (assuming local homogeneity over those distances). To find the inclination angle, $\alpha$, the max $R_{\xi\xi}$ was found for each wall normal location and a line was fit to these points. This process is illustrated in figure~\ref{fig:InclinAngle}(a,c). The angle between this line and the horizontal is the angle of inclination, similar in approach to that of \cite{TalluruChauhan-2020} (who did this using temporal, point measurements). 
\par
Figure \ref{fig:InclinAngle}(b,d) shows the variation in the inclination angle, $\alpha$, with increasing downstream distance for log- and wake-injection cases.
The inclination angle for the log-region injection decreases downstream showing a change of roughly $5\degree$ in inclination angle across the \RevisionText{measurement domain.} 
This is likely due to the strong straining and rotation that the scalar plume experiences early on as it is transported away from the wall due to the turbulent mechanisms.
Figure \ref{fig:InclinAngle}(d) shows that, for wake injection, the inclination angle is roughly constant. 
This can be associated with the smaller shear experienced by the plume when dispersed within the wake-region. 

\cite{TalluruChauhan-2020} calculated and discussed this inclination angle using temporal measurements with two synchronized point sensors. From their measurements, they observed an inclination angle ranging from $29\degree-31\degree$ depending on the injection height. \RevisionText{Similar two-point scalar correlations using scalar field measurements were also performed by \cite{Miller-2005} for various downstream distances. They observed that the correlation contours start isotropic close to the injection location (i.e. $\alpha\approx 90^\circ$) and then increasingly incline with downstream distance (i.e. decreasing $\alpha$). These trends are similar to that observed in the current work. The exact streamwise variation of this angle close to a point-source (i.e. Stage-1 and Stage-2 mixing) and the dependence on the boundary layer and injection/injector parameters are not studied in sufficient detail. It appears that these scalar concentrations approach the inclination angles comparable to that of $R_{uu}$ at very large distances from the injector \citep[Stage-3]{TavoularisCorrsin-1981}.
Close to the injector, we suspect that this angle is a strong function of the advection velocity, distance, and the mean vorticity experienced by the plume. 
} \RevisionText{From this standpoint, the relative size of the plume compared to the boundary layer thickness ($\sigma_y/\delta$) becomes relevant, which is directly proportional to the injector diameter. This is significantly different between the current work ($d_i/\delta=0.026$) and \citet{TalluruChauhan-2020} ($d_i/\delta\approx 0.005$) which likely accounts for the differences between the two works. 
However, it is still unclear how the plume parameters ($x/\delta$, $y/\delta$, $\sigma_y/\delta$ etc.) influence this inclination angle and further studies are necessary to understand these relationships.}
\section{Scalar Intermittency and `Blob'-like Behavior}
\label{sec:blob-statistics}
The turbulent mixing of the plume can be described phenomenologically in three stages as seen in figure \ref{fig:BlobSchematicV2}. The first stage lasts for a relatively short distance where the plume forms a laminar stream in the boundary layer. The shear forces and turbulence within the boundary layer destabilizes the plume which quickly transitions into the intermediate stage (stage 2). This stage of mixing is characterized by discrete parcels of concentrated fluid, hereafter referred to as `blobs' due to their appearance in the plane of measurement (note that these streaks extend continuously out-of-plane). After the blobs have formed, they may breakup into smaller blobs, be strained into elongated streaks, or diffuse and mix as small-scale turbulent motions ($\Order{< D_p}$ where $D_p$ is the local plume diameter). As this process continues over large downstream distances, the scalar field expands throughout the boundary layer, becomes less intermittent and forms a Continuous Scalar Field (CSF) (stage 3). 
\par
\begin{figure}
    \centering
    \includegraphics[width=1.0\textwidth]{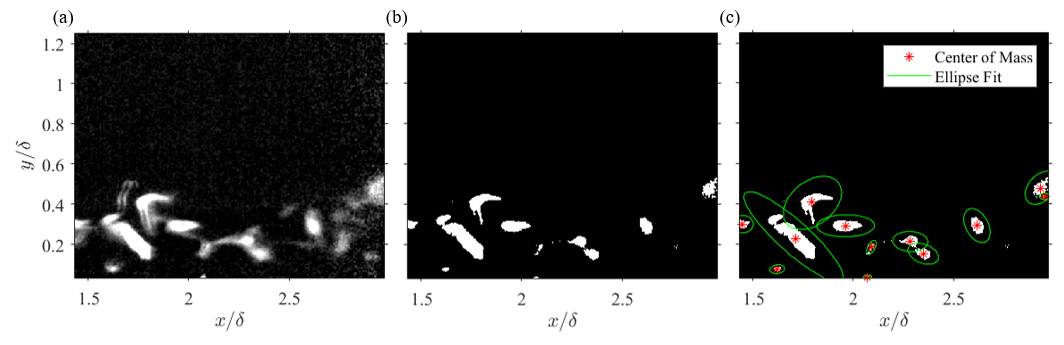}
    \caption{Example of processing to identify and extract blob properties: (a)~corrected image, (b)~mask applied (c)~area filter applied and center of mass location identified}
    \label{fig:LogExProc}
\end{figure}

The entirety of the plume evolution (stages 1-3 in figure \ref{fig:BlobSchematicV2}) has been investigated with experimental and theoretical endeavors. 
However, the intermediate or transitional stage has been difficult to characterize and lacks the breadth of investigation afforded to the initial and late stages of plume evolution. The current study will focus on this intermediate stage and attempt to characterize it by describing the statistical characteristics of blobs \RevisionText{captured in the instantaneous fields}. To this end, roughly $2,670$ mixture fraction fields were captured for each injection location and \RevisionText{the data were} extracted from the blobs in each frame. Figure~\ref{fig:LogExProc} details this process for an example mixture fraction field (figure~\ref{fig:LogExProc}a). 
This image is thresholded to create a mask (figure~\ref{fig:LogExProc}b) using a threshold of {$\xi_{th}=0.225$}. \RevisionText{The choice of this threshold was guided with the goal of detecting large regions of high-concentration blobs of marked scalar. We intentionally excluded (1)~small regions of high concentration (mostly from speckle noise) and (2)~large regions of diffused/well-mixed marked scalar.} \RevisionText{$\xi_{th}$ was defined quantitatively to be the 90th percentile of the distribution of $\xi$ values above the noise floor.}
\RevisionText{Applying this thresholded} provides the discrete boundaries of the blobs of high concentration, and the closed regions are then catalogued. 
Finally, an area filter is applied where any blob with an area less than a minimal threshold, $A_{th}=15\,px^2=1.58 \times 10^{-4}\,\delta^2$ is disregarded, effectively serving as a noise filter to exclude random bright spots. For each blob, the area ($\tilde{A}$), average mixture fraction ($\tilde{\xi}$), aspect ratio ($\widetilde{AR}$), inclination angle with respect to the wall ($\tilde{\phi}$), and center of mass ($\tilde{x}_{cm},\tilde{y}_{cm}$, marked in figure~\ref{fig:LogExProc}c) were extracted (note that we use $\tilde{(\cdot)}$ to represent blob properties). A total of 75,866 blobs were captured for both injection locations, which was reduced to 42,019 blobs after the area thresholding. 
\par
Using this catalog of blobs, we first study the concentration and coordinate data of the blobs to validate the meandering plume model proposed by \citep{Gifford-1959,FackrellRobins-1982, FackrellRobins-1982a}, and expanded on by \citet{TalluruChauhan-2020}.
Secondly, the relationship between blob area, $\tilde{A}$ and density, $\tilde{\rho}=\tilde{\xi}/\tilde{A}$ is investigated through the joint pdfs. The proposed relationship between area and density provides insight into blob dynamics in the intermediate region. Finally, the inclination angle and aspect ratio are investigated to provide insight into the physical shape and orientation of the blobs. 
\subsection{Scaled Probability of Concentration at Different Injection Points}
\label{subsec:ScaledProb}
\citet{Gifford-1959} presented the meandering plume model (also discussed in \cite{MarroNironiEtAl-2015}) in which the plume mixing occurs through two simultaneous phenomena captured via their PDFs. The first component of this model is the meandering of the center of mass ($p_m$ in equation \ref{eq:MeandPlume}) of the plume where large scale structures transport the entire plume along the wall-normal and span-wise directions. The second component of the theory encompasses the local dissipation of the plume through small scale turbulence and random molecular motion ($p_{cr}$ in equation \ref{eq:MeandPlume}). \begin{equation}
    \label{eq:MeandPlume}
    p(\xi;x,y,z)=\int_{0}^{\infty}\int_{-\infty}^{\infty}p_{cr}(\xi,x,y,z;y_m,z_m)p_m(y_m,z_m; x) dy_m dz_m
\end{equation}
With $\xi$ denoting the mixture fraction, $x,y,z$ denoting the stream-wise, wall-normal, and span-wise position respectively, and $y_m$ and $z_m$ denoting the $y$ and $z$ locations of the center of mass.
Critically, the model assumes that the two processes are independent of each other, an approximation that holds in high-$Re$ boundary layers when the plume diameter $D_p \ll \delta$ (i.e. $p_{cr} = fn(\xi, x, y-y_m,z-z_m\,\textrm{only})$).
This implies that, in the frame of reference of the plume's center of mass, its statistical characteristics (such as distributions of concentration, inclination angle, AR etc.) for a given $x$ will be independent of $y$ and $z$.
In other words, if the plume meanders to a given location, then its statistical characteristics should be identical to other wall-normal locations (though the probability of this meandering, $p_m$ will be different). 

\begin{figure}
    \centering
    \includegraphics[width=1.0\textwidth]{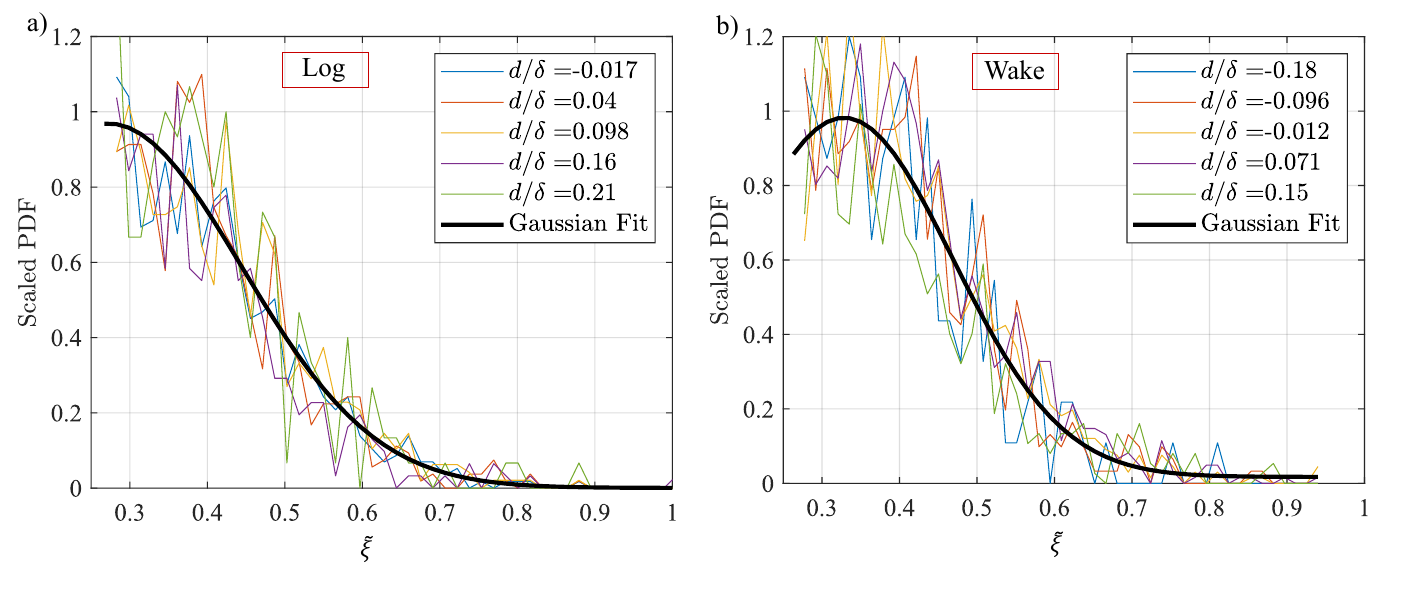}
    \caption{Normalized pdf of $\tilde{\xi}$ at different wall-normal distances from the injection location for (a)~log- and (b)~wake-injection cases.}
    \label{fig:ScaledInjLoc}
\end{figure}
We evaluate this condition via the pdf of the average mixture fraction, $\tilde{\xi}$ for both the \RevisionText{log- and wake-injection cases.}
Figure \ref{fig:ScaledInjLoc} shows the scaled (by maximum value) blob concentration probabilities at different wall-normal distances from the injection point. The data suggests that while the average total concentration decreases further from the injection point, the distribution of blob concentration at an arbitrary distances from the injection point seem to be identical. Phenomenologically this suggests that blobs of different concentrations indiscriminately meander from the injection line such that the statistical distribution of the concentration of the blobs at any wall-normal distance form the injection line are identical. 
\RevisionText{This supports the statistical independence of the meandering and dispersion mechanisms described in the meandering plume model. However, further studies, ideally with a well converged dataset and a range of injection locations, are necessary to confidently establish the statistical independence between wall-normal location and distribution of blob mixture fraction.} 

\subsection{Relationship Between Blob Density and Area}
Using the blob database, we try to understand the mechanisms by which the discrete blobs `breakup' and `disperse' into a continuous field, as was shown in figure~\ref{fig:BlobSchematicV2}. Both these processes change the area of the scalar blobs observed. An ideal `breakup' occurs due to locally diverging vector topology (saddle-like topologies in plane), and preserves the average concentration of scalar in the blob (i.e. $\tilde{\xi}$). The `dispersion' occurs due to micromixing by small-scale local turbulence ($l\ll D_p$), and preserves the amount of substance in the blob (i.e. $\tilde{\xi}\tilde{A}$). Even though the transport is a three dimensional phenomenon, these two processes can be statistically visualized using the distribution of blob areas and concentrations in the symmetry plane.
\begin{figure}
    \centering
    \includegraphics[width=1.0\textwidth]{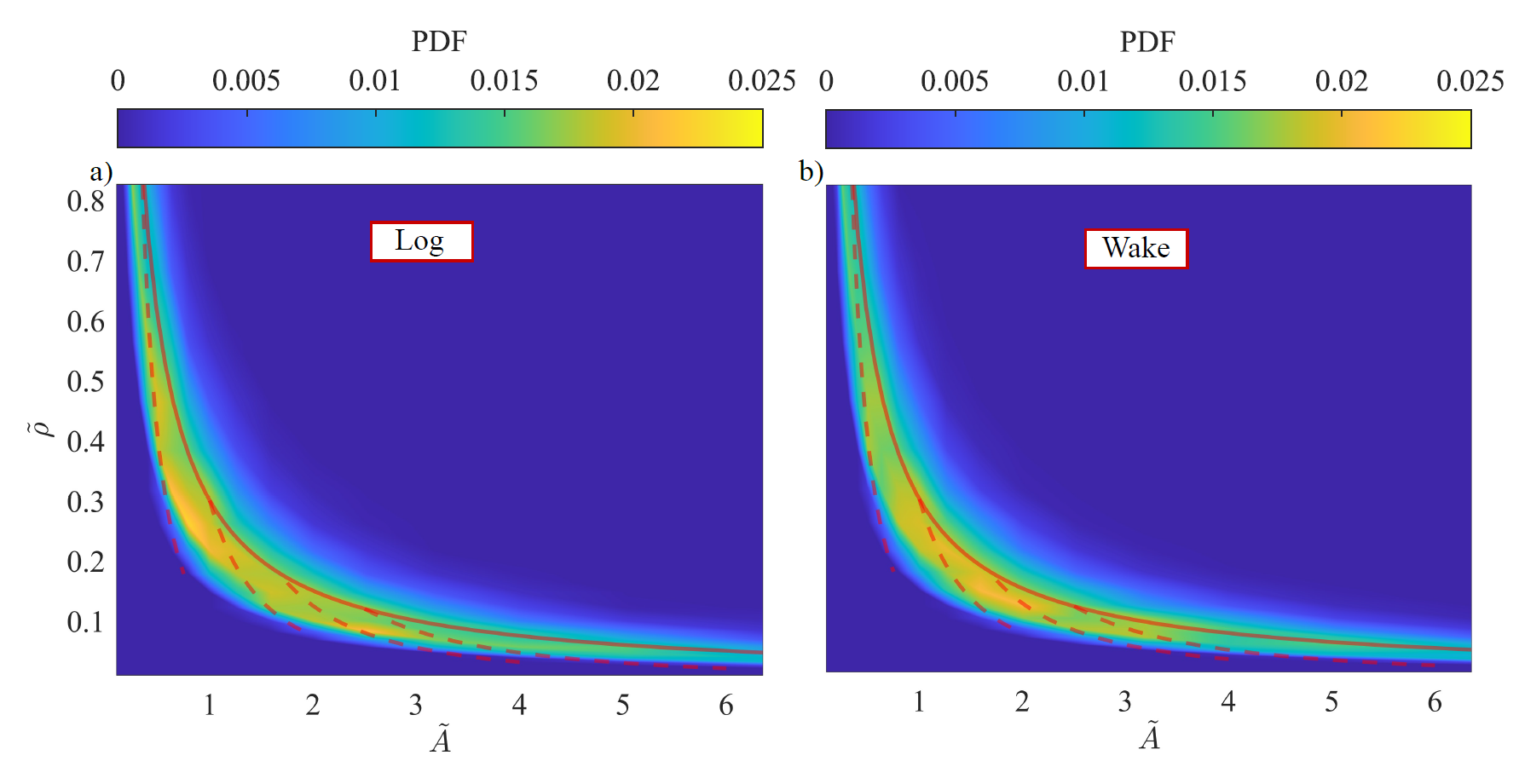}
    \caption{Joint pdf of $\tilde{\rho}$ and area $\tilde{A}$ for (a)~log- and (b)~wake injection cases.}
    \label{fig:RhoAJPDF}
\end{figure}

Figure \ref{fig:RhoAJPDF} shows the JPDF of scalar density (defined as $\tilde{\rho}=\tilde{\xi}/\tilde{A}$) and blob areas ($\tilde{A}$) for the log- and wake- injection cases. Both cases show a distribution of $\tilde{\rho}$ and $\tilde{A}$. For a pure breakup event, where a blob breaks up into smaller blobs, the area would decrease but the average mixture fraction $\tilde{\xi}$ would be conserved. A successive progression of this process will result in a distribution of blobs cascading up a curve of $\tilde{\rho}=k_1/\tilde{A}$ ($k_1$ is an arbitrary constant depending on the initial concentration). The solid red line represents an example of this with $k_1=0.3$ ($k_1=0.25$ provides the best fit assuming only blob breakup). On the other hand, should the blob change in physical size due to diffusion, the area changes to conserve the material, $\tilde{m}=\tilde{\rho}\tilde{A}^2$. This results in a cascade along the line $\tilde{\rho} = k_2/\tilde{A}^2$ ($k_2$ is a constant depending on initial blob mass). The dashed lines in \ref{fig:RhoAJPDF} represent this diffusion cascade for several $k_2$ values illustrating how dispersion causes the blob evolution to stray from the purely breakup case (solid line). 
Given \RevisionText{an} initial distribution of blobs with large area and low $\Tilde{\rho}$) (bottom right of figure \ref{fig:RhoAJPDF}), the evolution of these blobs along a purely breakup trajectory would account for the general shape of the JPDF \RevisionText{including some of the} spread perpendicular to the $\Tilde{\rho}=k/\Tilde{A}$ line. The spread of the JPDF perpendicular to this line of $\Tilde{\rho}=0.25/\Tilde{A}$ is caused from both the variation in initial blob size and the dispersion along the dashed lines (diffusion). 
The strong trends of the JPDF along the breakup line, and weak spread perpendicular to it, indicate that blob breakup is the primary mechanism of plume evolution in the stream-wise extent considered. However, as the plume disperses within the boundary layer further downstream and the scalar spectrum develops length scales $\Order{\delta}$, dispersion is expected to play a more significant role, limiting the utility of this discrete blob approach.
\par
\RevisionText{The evolution of the blob area and mean concentration as a function of stream-wise location can also be investigated to determine the dominant mechanism of plume evolution. If the plume evolution at this stage is breakup dominated, then the average area should decrease as the stream-wise position increases. Since breakup does not change the mixture fraction $\tilde{\xi}$, the mixture fraction $\tilde{\xi}$ should stay relatively constant. Alternatively, if the plume evolution is dispersion dominated, the opposite trends are expected. The average area of the blobs should increase with streamwise position, with an associated decrease in the mean mixture fraction.}   
\begin{figure}
    \centering
    \includegraphics[width=1.0\linewidth]{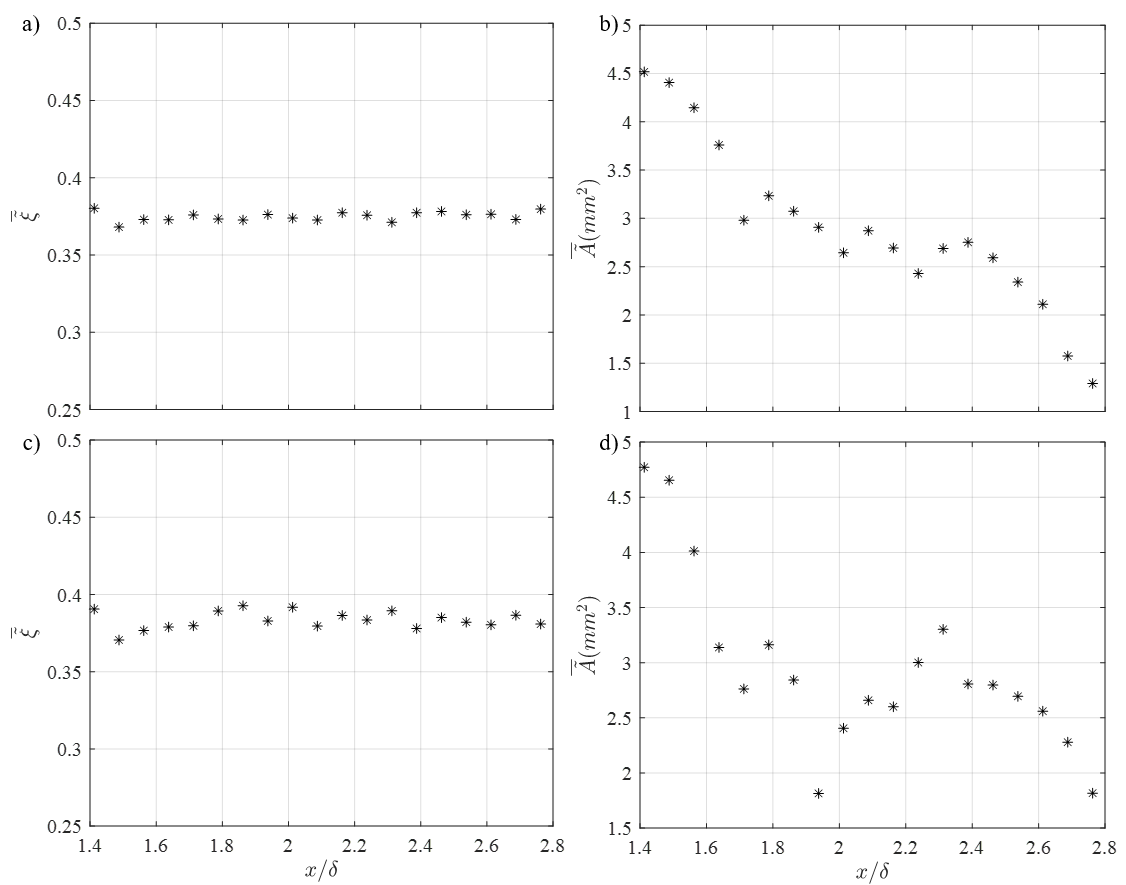}
    \caption{Mixture fraction of the blobs (a,c), and area of the blobs (b,d) at different stream-wise locations for log (a,b) and wake (c,d) injection}
    \label{fig:XiA_Streamwise}
\end{figure}
\RevisionText{The trends of blob mixture fraction and area as a function of streamwise location can be seen in figure \ref{fig:XiA_Streamwise}. 
The evolution of the average blob area ($\overline{\tilde{A}}$) shows a marked decrease across the FOV. The intermittent peaks in the average blob area (figure \ref{fig:XiA_Streamwise}(b,d)) are likely caused by the lack of convergence. The average mixture fraction of the blobs ($\overline{\tilde{\xi}}$) stays relatively constant ($\approx\pm3\%$) across the FOV. 
The marked decrease in area with downstream location seen in figure \ref{fig:XiA_Streamwise}(b,d), and the constant value of mixture fraction across the stream-wise locations (figure \ref{fig:XiA_Streamwise}(a,c)) supports the breakup dominated plume evolution previously discussed.}
\subsection{Aspect Ratio and Inclination Angle of Blobs}
\begin{figure}
    \centering
    \includegraphics[width=1.0\textwidth]{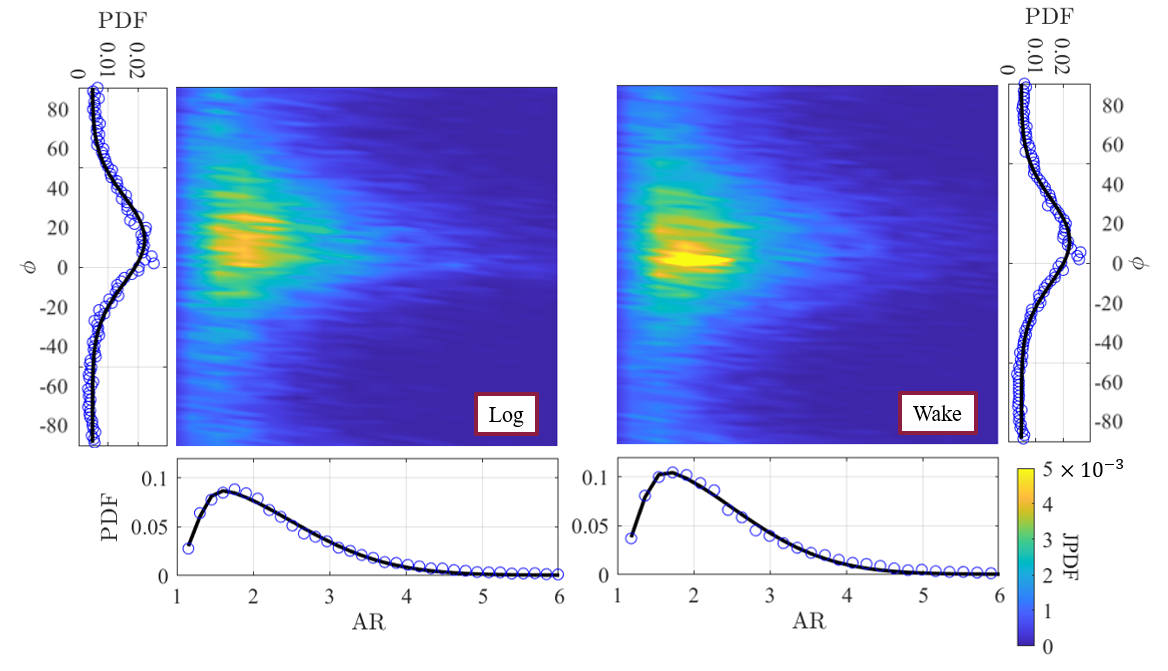}
    \caption{Joint PDF and PDFs of aspect ratio ($\widetilde{AR}$) and inclination angle ($\tilde{\phi}$) for log- and wake- injection cases}
    \label{fig:PhiAR}
\end{figure}

\begin{wraptable}{r}{0.5\textwidth}
    \caption{Best fit distribution constants for the log- \textit{(left)} and wake- (\textit{right}) injection cases}
    \label{tab:phiARConstants}
    \begin{tabular}{c||c}
    Inclination Angle ($\tilde{\phi}$) & Aspect Ratio ($\widetilde{AR}$)\\
    \begin{tabular}{|c|c|c|} \hline - & $\mu$ & $\sigma$\\ \hline log & 12.5 & 21.3\\ \hline wake & 10.1 & 21.4\\ \hline \end{tabular} &
    \begin{tabular}{|c|c|c|} \hline - & $\mu$ & $\sigma$\\ \hline log & 1.17 & 1.22\\ \hline wake & 1.18 & 1.18\\ \hline                  \end{tabular}
    \end{tabular}
    \end{wraptable}
Besides the size and \RevisionText{mixture fraction}n, the scalar blobs are also elongated and inclined with the horizontal at a certain angle. These characteristics were also investigated using the ensemble.
\RevisionText{For each blob, an ellipse was fitted to the outline (figure~\ref{fig:LogExProc}b) with an equivalent area moment of inertia about the major and minor axes \citep{Matlab-regionprops-2024}.}
\RevisionText{The $\widetilde{AR}$ can then be calculated as the ratio of the major to minor axes ($a/b$), and $\phi$ as the counterclockwise angle between the horizontal and the major axis of the fitted ellipse.} Its important to note that the blobs rarely formed perfect ellipses; however, using this approach provides a consistent method to calculate these two quantities which provide insight into the statistical nature of the blobs' evolution. Inclination angle provides insight into how the blobs tend to rotate and how they are formed, while the AR gives insight into how the blobs tend to be stretched and elongated. 
Clear trends can be seen in figure~\ref{fig:PhiAR} for the $\widetilde{AR}$ and inclination angle $\tilde{\rho}$ at both injection locations. A normal distribution was fitted to the inclination angle, which provide a good fit to the data. A skewed normal distribution was used to fit the aspect ratio providing a good estimate of the observed measurements.

The constants in table~\ref{tab:phiARConstants} summarize the various constants for the distribution (mean, $\mu$ and variance, $\sigma$). It can be seen that there is a clear positive preference for the inclination angle of the blobs in both injection regions. The log injection has a slightly larger inclination angle at $ 12.5 \degree$ compared to $10\degree$ in the wake injection. This is due to larger shear that the blobs experience as they evolve from the injection height. The standard deviation for both regions are statistically identical. For the aspect ratio, both regions have a peak around 1.6-1.7 with a steep decline in pdf afterwards. Both regions show a strong preference for an elongated $(AR>1)$ shape, with the majority of the blobs having an aspect ratio $1<AR<3$. The only notable difference between the AR distribution for the log- and wake- injection is the wake distribution in AR is flatter than the log region, likely due to higher mean shear experienced by the log-injected plume and leading to a larger variation in aspect ratios. 
\section{Spatial coherence in plume intermittency} \label{sec:coherent-scalar-arrangement}
\subsection{Instantaneous observations}
\label{sec:instantaneous-coherence}
\begin{figure}
    \centering
    \includegraphics[width=\textwidth]{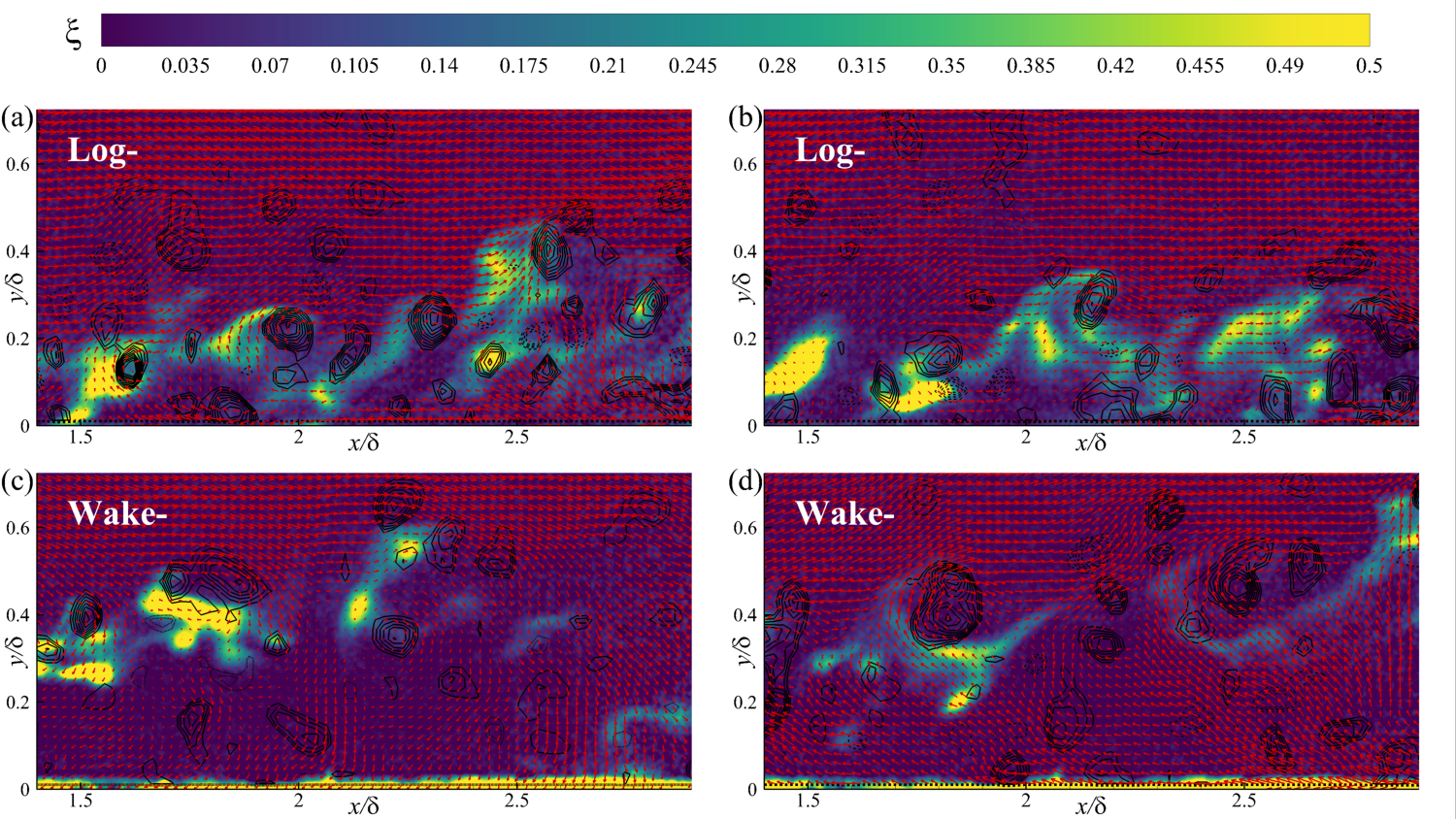}
    \caption{Instantaneous snapshots showing scalar plume together with contours of swirling strength. Vectors are shown in a frame moving at $\approx 0.88 U_\infty$}
    \label{fig:instantaneous-coherence}
\end{figure}
With the previous observations indicating that the scalar plume is highly intermittent and dispersed into discrete \textit{blobs}, we can focus on the role of specific turbulence structures in developing the spatial trends. In other words, a continuous injection of the passive scalar is reorganized into intermittent and concentrated pockets by the action of turbulent structures of similar scale. Figure~\ref{fig:instantaneous-coherence} shows instantaneous snapshots of this organization, in an arbitrary frame of reference moving with the fluid for the two injection configurations. Also shown are the contours of swirling strength, $\lambda_{ci}$, that represent the local rotational motion of the fluid elements due to strong vortices (the swirling motion is not immediately evident in the vector field due to advection). The swirling strength, $\lambda_{ci}$ is defined as the imaginary part of the complex eigen value of the velocity gradient tensor, and local maxima in its value indicate the location of a vortex element \citep{ChakrabortyBalachandarEtAl-2005}. The current experiments are only capable of capturing the vortex elements sliced perpendicularly by the measurement plane, and the sign of the vorticity is assigned to identify the `prograde' ($\lambda_{ci}\leq0$, clockwise) and `retrograde' ($\lambda_{ci}>0$, anti-clockwise) vortices \citep{AdrianChristensenEtAl-2000a}.

A qualitative organization of the scalar plume relative to the vortex cores is observed in figure~\ref{fig:instantaneous-coherence}(a,b) -- a trend that occurs around 25\% of the vector fields. This trend is particularly dominant if the fields are conditioned on a scalar blob occurring far away from the injection location ($y-y_{i}>2\sigma_y$ ). It can be seen that, a discrete scalar blob occurs preferentially between the vortex cores, and at an angle inclined $\approx\,15\degree-18\degree$. Additionally, multiple vortex cores are observed to be aligned along this line with the distinct signature of the coherent vortex packet \citep{AdrianMeinhartEtAl-2000, ChristensenAdrian-2001, Adrian-2007}. This alternating inclined scalar-blob--vortex-core organization was less prevalent when the injection was in the wake region, as illustrated in figure~\ref{fig:instantaneous-coherence}(c,d). The scalar plume was still dispersed and intermittent, and the discrete scalar blobs were mostly attached to a dominant vortex core. However, coherent interlacing of scalar blobs with an inclined set of coherent vortices extending all the way to the wall was not observed for wake-injection. This indicates that the action of the vortices (possibly from the detached packets) on the scalar plume is still the dominant cause of the scalar plume intermittency, however without the spatial coherence.
Finally, it must be noted while interpreting these snapshots that we are only measuring a single plane, and that we are limited to observing these coherent trends only when the coherent vortex packet is aligned with the measurement/injection plane.

\subsection{Statistical footprint with coherent vortex packets}
\label{sec:statistical-footprint-with-cvp}
\subsubsection{Conditional structure for log injection}
\begin{figure}
    \centering
    \includegraphics[width=\textwidth]{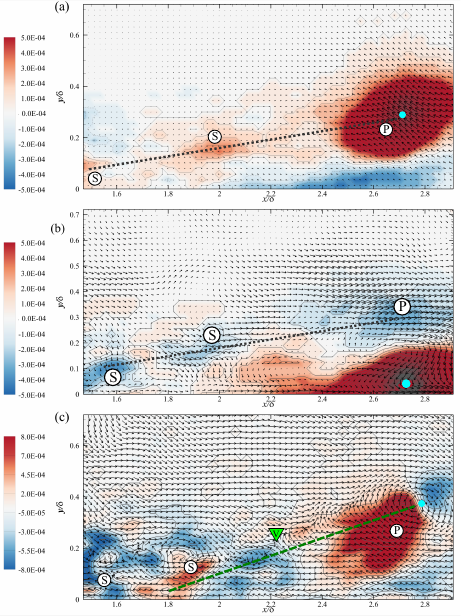}
    \caption{Conditional fields of scalar concentration perturbation and relative fluid velocity. (a,b) are conditioned on a positive scalar fluctuation and vectors show the velocity field, (c)~is conditioned on a clockwise swirl-event and vectors show velocity direction. \textit{Blue} circles indicate the reference locations for each event.}
    \label{fig:lse-cvp-log-injection-p1}
\end{figure}

Based on these qualitative observations, we can devise a statistical approach to investigate the dominance of these trends via conditional fields. However, since the conditional averages require a large quantum of data, we estimate the same using the linear stochastic estimation approach previously employed to observe similar spatial trends \citep{AdrianJonesEtAl-1989a, Adrian-1994}. The conditional field of a quantity, $Q (\overline{x})$ conditioned on the occurrence of a second quantity (`\textit{event}'), $Q_E(\bar{x}_{r})$ at reference location, $\bar{x}_{r}$ is given as,
\begin{align}
    \langle Q(\bar{x})|Q_{E}(\bar{x}_{r})\rangle \approx
    \frac{\langle Q(\bar{x})Q_{E}(\bar{x}_{r}) \rangle}{\langle Q_{E} Q_{E}\rangle} Q_{E}(\bar{x}_{r})
\end{align}
That is, the two-point correlation function embodies the information of the conditional field to the first order, when conditioned on a single scalar quantity.
Since the aforementioned coherent reorganization was found consistently upstream when the scalar plume is transported far away from the injection location, we first choose this as the reference event, i.e. $\bar{x}_{r}=(2.75\delta, 0.3\delta \equiv h_i + 1\sigma_y)$ to estimate the concentration ($\langle \xi'(\bar{x})|\xi'(\bar{x}_{r})\rangle$) and velocity fields $\langle \bar{u}(\bar{x})|\xi'(\bar{x}_{r})\rangle$. This is shown in figure~\ref{fig:lse-cvp-log-injection-p1}a, that shows the conditional concentration fluctuation and velocity fields in the FOV. A few aspects are immediately obvious from the conditional field. The correlation near the wall at $x\approx x_{r}$ reaches negative values, which implies that a high concentration event away from the wall is accompanied by a low concentration close to the wall at the same stream-wise location. This is expected for a conserved, meandering scalar. The conditional velocity fields around the reference location, $\bar{x}_{r}$, also exhibit a positive $v'$ and negative $u'$ (an `ejection/Q2' event). This is entirely consistent with the turbulent flux at $\bar{x}_{r}$, discussed earlier in section~\ref{sec:turbulent-flux}. The most striking observation in the conditional scalar field is that a high-concentration event at $\bar{x}_{r}$ (we refer to this region of coherence as `primary blob, \texttt{P}') is accompanied by two `secondary blobs' of positive concentration fluctuation, upstream and closer to the wall (marked by \texttt{S} in figure~\ref{fig:lse-cvp-log-injection-p1}a). The line joining these discrete, high concentration regions is inclined at $\approx 10\degree$, consistent with the inclination angle of coherent vortex packets \citep{AdrianMeinhartEtAl-2000, ChristensenAdrian-2001, Adrian-2007}. The robustness of these discrete secondary blobs is also observed in a statistically independent, but complementary measure in figure~\ref{fig:lse-cvp-log-injection-p1}b. 
Here the correlations for a positive concentration fluctuation in a region close to the wall, i.e. $\langle \xi'(\bar{x})|\xi'(\bar{x}_{r})\rangle$ and $\langle \bar{u}(\bar{x})|\xi'(\bar{x}_{r})\rangle$ with $\bar{x}_{r}=(2.75\delta, 0.04\delta)$, are shown. Not only does the anti-correlation at the outer location corresponding the primary concentration blob at $\bar{x}=(2.75\delta, 0.3\delta)$ appears (as expected), but we also see the anti-correlations blobs at the secondary locations [$\bar{x}\approx(1.95\delta, 0.18\delta)$ and $(1.6\delta, 0.1\delta)$] at exactly the same locations. This relative correlation structure by two independent measures demonstrates the robustness of this spatial organization. Further, these observations are insensitive to the exact location of the $\bar{x}_r$ chosen, as long as it is in the vicinity of the primary blob seen here. 
\RevisionText{Finally, the conditional structure in figure~\ref{fig:lse-cvp-log-injection-p1}b indicates a consistent Q3 event (i.e. $u'<0, v'<0$) that is unusual to observe in a boundary layer. This structure appears in the relative frame as a pair of counter-rotating vortices, and demonstrate that the scalar below the injection location occurs predominantly in the low-momentum regions, indicating a strong relation with a UMZ. This observation was insensitive to the choice of $x_r$ for all choices of $y_r<y_{inj}$.}

\begin{figure}
    \centering
    \includegraphics[width=0.8\textwidth]{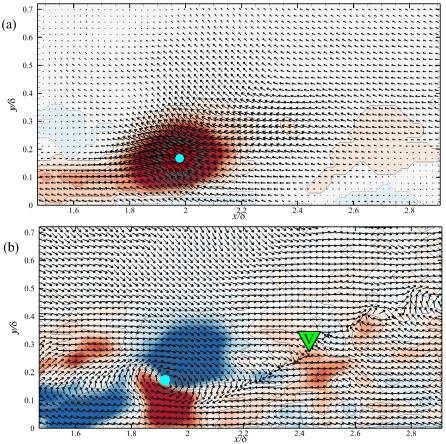}
    \caption{Same as figure~\ref{fig:lse-cvp-log-injection-p1}(a,c), but for a different reference location.}
    \label{fig:lse-cvp-log-injection-p2}
\end{figure}

If we were to hypothesize that this discrete spatial coherence in scalar concentration occurs due to the action of a coherent vortex packet, as was seen in figure~\ref{fig:instantaneous-coherence}, the same should be observable in the conditional estimates based on the occurrence of a swirl event, i.e. $\langle \xi'(\bar{x})|\lambda_{ci}(\bar{x}_{r})\rangle$. 
A swirl event for $y_{r}>y_{inj}$ was found to generally have a positive mixture fraction, $\xi'$ below-upstream and a negative concentration above-downstream of the reference location. This velocity-concentration field is consistent with that observed for a primary blob in figure~\ref{fig:lse-cvp-log-injection-p1}a. Thus to induce the primary scalar blob at the same location, i.e. $\bar{x}=(2.7\delta, 0.3\delta)$, a swirl event at $\bar{x}_{r}\approx(2.77\delta, 0.37\delta)$ is considered, as shown in figure~\ref{fig:lse-cvp-log-injection-p1}c. 
The typical signatures of a coherent vortex packet \citep{AdrianMeinhartEtAl-2000, ChristensenAdrian-2001,Adrian-2007} in the conditional velocity field ($\langle \hat{u}(\bar{x})|\lambda_{ci}(\bar{x}_{r}=(2.77\delta, 0.37\delta))\rangle$) are reproduced accurately (for clarity, velocity direction $\langle \hat{u}|\lambda\rangle = \langle \bar{u}|\lambda_{ci}\rangle/|\langle \bar{u}|\lambda\rangle|$, with unit magnitude, is shown in figure~\ref{fig:lse-cvp-log-injection-p1}c). Specifically, the primary hairpin head, and two `younger', secondary hairpins appear with an inclined shear flow between them. Remarkably, the conditional scalar field shows the three positive peaks associated with the primary and secondary blobs at their respective locations identified in figure~\ref{fig:lse-cvp-log-injection-p1}a. 
This shows direct correlation between the coherent vortex packets and the scalar concentration blobs, as was hypothesized from instantaneous observations in figure~\ref{fig:instantaneous-coherence}(a,b). Additionally, the scalar blobs seem to be isolated from each other with influx events from the upstream vortex and a VITA event (saddle-point-like vector topology, marked `\texttt{V}', \cite{Adrian-2007}) that serves as topological barriers for the passive scalar. This provides a dynamical mechanism based on the coherent vortex packets for origination and development of discrete coherent scalar packets.

{\color{Rev1}
First, it must be noted that the evidence presented in Figures~\ref{fig:lse-cvp-log-injection-p1}(a, b) does not, by itself, substantiate the conclusion that the vortex packets are responsible for the coherent scalar arrangement. They only demonstrate that the scalar is distributed in discrete packets along an inclined line. This arrangement was found to be very robust and insensitive to the choice of $x_{r}$ (for a similar $y_{r}$). The relation between the coherent vortices and the relative scalar field is seen only when a conditional swirl event is observed (figure~\ref{fig:lse-cvp-log-injection-p1}c). As we present it here, the equivalence in the two observations is hypothesized only due to the compelling similarities in the two independent measures, and additional trends discussed in the appendix. Additionally, we do not notice secondary swirl in the velocity fields conditioned on concentration event in figures~\ref{fig:lse-cvp-log-injection-p1}(a,b). This could be a combination of following factors: 
\begin{itemize}
    \item The advection of the high concentration blob will dominate the velocity features observed in the conditional average, rather than the swirl and the long-range spatial coherence.
    \item We expect a high concentration blob to span a larger {spanwise} extent than the swirl field. This results in averaging over velocity features that are not in the symmetry plane, and in decreasing the long-range correlation of coherent vortices (dominant only in the symmetry plane). The swirl condition (figures \ref{fig:lse-cvp-log-injection-p1}c) does not suffer from this as it decays faster in spanwise direction (due to hairpin shape) and the misaligned cases do not contribute to the correlation. Figure~\ref{fig:lse-cvp-log-injection-p1}a represents an average that is agnostic to this distinction. We demonstrate via analysis presented in the appendix that about $50\%$ of the concentration events occur coupled with a swirl event. A direct average conditioned on both events does elucidate a weak secondary swirl (in figure~\ref{fig:direct-conditional-averaging}) coupled with a secondary concentration blob.
\end{itemize}
}

In addition to the primary blob as reference, we investigate the conditional fields based on the occurrence of near-wall secondary blob, i.e. a discrete concentration event at a location $\bar{x}_{r}=(1.97\delta, 0.17\delta)$. This is shown in figure~\ref{fig:lse-cvp-log-injection-p2}a. No correlated downstream--outer peak in concentration representative of the earlier primary blob is observed, indicating that the concentration fluctuations at this location are not necessarily correlated with a coherent arrangement of concentration blobs (as was seen in figure~\ref{fig:lse-cvp-log-injection-p1}a). This indicates that there are other mechanisms at play beyond that of a large attached, coherent vortex packet that discretize the injected stream of passive scalar into pockets at this location. In other words, an occurrence of a discrete blob at an injection location is not necessarily an indication of a coherent vortex packet. This is entirely consistent with qualitative observations made from the instantaneous snapshots where, on more occasions than not, a discrete scalar blob at the injection location is not accompanied by an inclined arrangement of coherent blobs. Finally, if we consider the occurrence of a vortex event ($\langle \xi'(\bar{x})|\lambda_{ci}(\bar{x}_{r})\rangle$) close to the wall instead of a concentration event ($\langle \xi'(\bar{x})|\xi'(\bar{x}_{r})\rangle$), we do see a corresponding coherent arrangement of discrete blobs along the inclined shear layer, as shown in figure~\ref{fig:lse-cvp-log-injection-p2}b. This definitively demonstrates the action of the coherent vortex packets in the discretization and transport of the passive scalar along the shear layer. Finally, it is worth noting that the observations noted here were insensitive to the exact values of $\bar{x}_r$ shown here, as long as they were in the same approximate regimes.

{\color{Rev1}

\begin{figure}
    \centering
    \includegraphics[width=\textwidth]{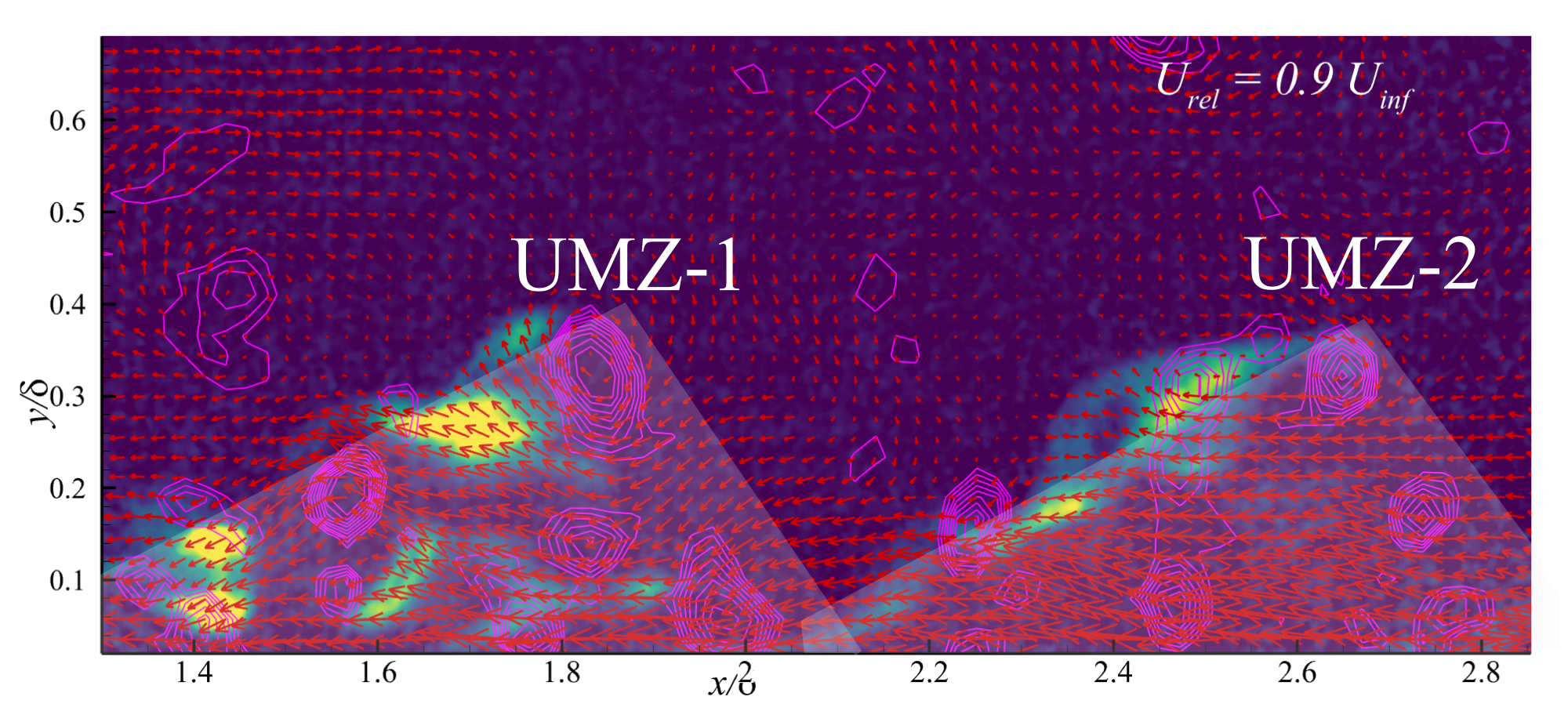}
    \caption{An illustrative snapshot qualitatively showing ramp-like UMZ features and relative arrangement of vortex packets and scalars. Velocity field shown relative to a moving frame of reference, $U_{rel}=0.9U_\infty$. Magenta contours represent contours of high $|\lambda_{ci}|$, and scalar contours are same as figure~\ref{fig:instantaneous-coherence}.}
    \label{fig:umz-features}
  \end{figure}
We can contextualize these observations with that observed in prior works that evaluated the large-scale structure associated with scalars \citep{AntoniaFulachier-1989,LaskariSaxton-FoxEtAl-2020,EismaWesterweelEtAl-2021,Metzger-2002}. Specifically, both the works established the conditional large-scale structure associated with `coolings' (equivalently the trailing edge of a large, low-momentum bulge inclined in streamwise direction). The works of \cite{AntoniaFulachier-1989} established that these structures have a saddle-like topology in local turbulence, and \citet{LaskariSaxton-FoxEtAl-2020} established the fine-scale structure of the same, demonstrating a smaller-scale vortex distribution. The role of these structures on a point-source injected scalar is perfectly demonstrated in the conditional fields (figure~\ref{fig:lse-cvp-log-injection-p1}), and the illustrative instantaneous field show in figure~\ref{fig:umz-features}.
This integrates the advection arguments of \citet{AntoniaFulachier-1989} and \citet{LaskariSaxton-FoxEtAl-2020}, the coherent vortex packets \cite{Adrian-2007}, and the observance of these as uniform momentum zones (UMZs). Specifically, the trailing edge of the UMZs represent the `cooling' events \citep{AntoniaFulachier-1989}, and the vortices arranged represent the finer structure observed in \citet{LaskariSaxton-FoxEtAl-2020}. Both these works use temperature as a proxy for low-momentum regions. The current work demonstrates that the injected scalar `\textit{hitch-hikes}' on the `cooling events' resulting in a large meander event. In the process, the saddle-like topology induces the spatial discretization observed both instantaneously (figure~\ref{fig:umz-features}) and stochastically (figure~\ref{fig:lse-cvp-log-injection-p1}c). 
}

\begin{figure}
    \centering
    \includegraphics[width = \textwidth]{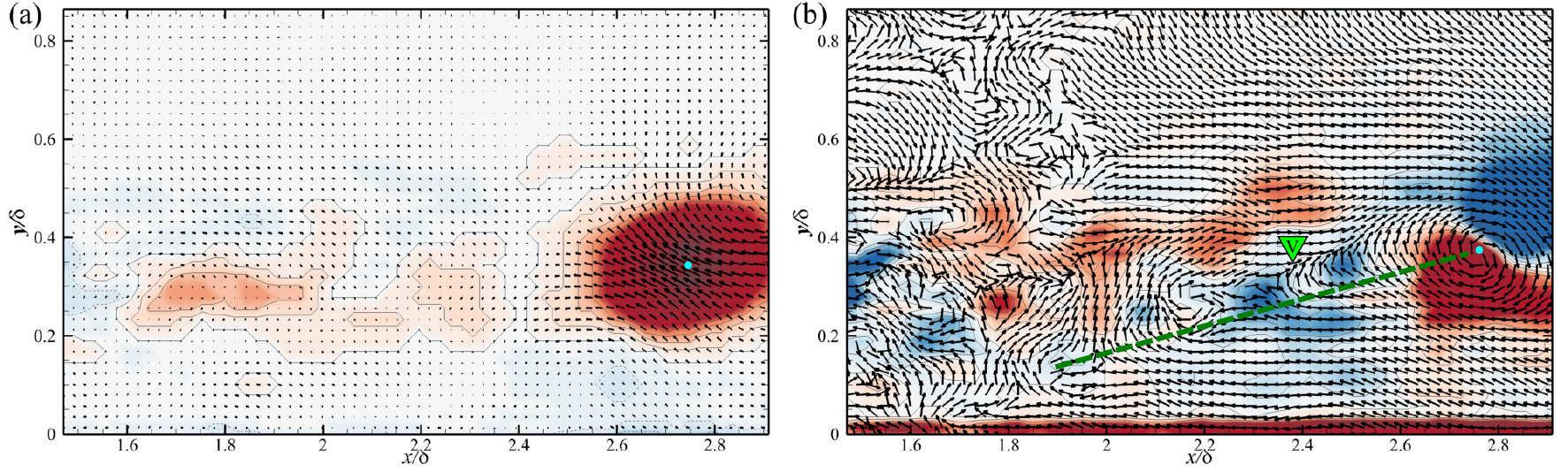}
    \caption{Same as figure~\ref{fig:lse-cvp-log-injection-p1}(a,c), but for wake injection.}
    \label{fig:lse-cvp-wake-injection}
\end{figure}

\subsubsection{Conditional structure for wake injection}
Similar conditional analysis can be performed when the scalar is injected into the wake region to see if any coherent trends of similar nature are observable. Figure~\ref{fig:lse-cvp-wake-injection}a shows the conditional fields, $\langle \xi'(\bar{x})|\xi(\bar{x}_{r})\rangle$ and $\langle \bar{u}(\bar{x})|\xi'(\bar{x}_{r})\rangle$, where a $\xi'>0$-event is observed slightly above the injection location ($\bar{x}_{r} = \left[2.75\delta, 0.35\delta\right]$). The meandering nature of the plume is evident, since the positive correlation at the reference location is accompanied by a negative correlation below the injection location. There also appears to be a secondary correlation peak up stream of the reference location along the injection line. This can be representative of mean separation distance of the discrete blobs. Further, recall from the instantaneous snapshots (figure~\ref{fig:instantaneous-coherence}) that discrete scalar blobs were consistently associated with vortices. This can also be observed in figure~\ref{fig:lse-cvp-wake-injection}b, where $\langle \xi'(\bar{x})|\lambda_{ci}(\bar{x}_{r})\rangle$ and $\langle \hat{u}(\bar{x})|\lambda_{ci}(\bar{x}_{r})\rangle$ fields are shown, with $\bar{x}_{r} = \left[2.75\delta, 0.35\delta \right]$. The swirl at $\bar{x}_{r}$ induces the ejection event $u'-v'$ seen in figure~\ref{fig:lse-cvp-wake-injection}a, and the relative $\xi'$-field is also shown. The occurrence of a discrete concentration blob is clearly seen here, with a scalar-excess region ($\xi'>0$) sandwiched between two scalar-deficit regions ($\xi'<0$). The relative organization of this $\xi'$ structure with the inclined shear-layer of the coherent vortex packet is also evident, with the VITA event (marked \texttt{V}) serving as a material barrier to the ejected plume. The \textit{sweep} event from the secondary vortex upstream and along the shear layer (marked \texttt{S}) appears to play a dominant role in breaking up of the plume by bringing in the scalar-deficit free-stream fluid. Beyond this primary blob, the secondary blob seen in figure~\ref{fig:lse-cvp-wake-injection}a does not appear in this vortex-conditioned field. However, the inclined shear layer that is characteristic of the coherent vortex packet does appear to direct the injected scalar plume away from the wall, while sustaining concentration gradient normal to it. Beyond these two aspects, i.e. the break up of the primary packet and the deflection of the upstream plume, the vortex packet does not seem to play a direct role in a coherent break-up and reorganization of discrete scalar plumes as was observed in the case of log-region injection.

From these observations, we can make a phenomenological model for the scalar plume dispersal for both the injection cases, as demonstrated in figure~\ref{fig:CVP-paradigm}. When the scalar plume is injected in the logarithmic region, the role of turbulence in breaking up the plume into discrete pockets of varying sizes (discussed previously in section~\ref{sec:blob-statistics}) can be viewed as a combination of two aspects -- (i)~an incoherent interaction of the plume with various vortical motions $\Order{1\,\sigma_y, 1\,\sigma_z}$, and (ii)~an interaction of the plume with a growing coherent vortex packet occurring near the wall. The former was not investigated in much detail in the current work. The interaction with the latter (itself a dominant dynamic process within a boundary layer) not only disperses the scalar plume in to discrete coherent pockets, but also transport these away from the wall into the wake region together with the oldest/largest hairpin-vortex. This appears to be the dominant mechanism by which the scalar plume  can reach the outer extents of the boundary layer ($y/\delta > 0.4$) in relatively high concentrations. 
This highlights the dominant role of the coherent vortex structures and a mechanistic transport mechanism of wall-normal transport of passive scalar close to the wall. On the other hand, when the scalar is injected in the wake region, the incoherent dispersal seems to be dominant mechanism by which the scalar plume breaks up. The interaction with the vortex packet only occurs at larger, older vortices that are detached from the wall. The spatial coherence has only a weak interaction with the upstream plume, which appears as a slight wall-normal deflection of the plume along the inclined shear-layer. This indicates that a purely statistical approach, such as a meandering plume model, can reasonably capture the scalar dispersal for a scalar injected in wake region. Finally, it is important to note that this coherent view point is only valid when the symmetry plane of the vortex packets is approximately aligned with the injection plane. They will have a role in spanwise meandering of the plume when there is a misalignment between the two planes, which is not investigated here. In other words, we are incorporating the influence of non-aligned vortex packets as the incoherent component in this paradigm. Measurements similar to these in streamwise--wall-parallel plane might be best suited for these studies.

\section{Discussion and conclusions} \label{sec:discussion-conclusions}
Implementing and validating the simultaneous planar velocity and quantitative mixture fraction measurements using synchronized PIV and Ac-PLIF, let us study the spatial aspects of the scalar plume evolution. Specifically, we focus on the region shortly downstream of the injection location ($1.5\le x/\delta \le 3$) when the scalar is injected isokinetically in logarithmic- and wake regions of the turbulent boundary layer. The plume behavior is dominated by strong meandering and exhibits high intermittency in this region \citep{FackrellRobins-1982a,CrimaldiWileyEtAl-2002}. We particularly focused on the stochastic description here, that can leave a footprint in the evolution of the scalar at far downstream locations.

\subsection{On the mean plume evolution and plume spread}
Given the novelty in the diagnostic approach, especially for flows in air, we first quantified the canonical nature of the experiments using the mean plume evolution and single-point statistics ($\overline{\xi'^2}, \overline{u'\xi'}, \overline{v'\xi'}$). This also established the baseline behavior. It was found that Gaussian plume model captures the mean evolution well, with its spreading behavior being reasonably captured by the analytical models \citep{NironiSalizzoniEtAl-2015}. Additionally, the meandering plume model \citep{Gifford-1959, MarroNironiEtAl-2015} was also found to quantitatively capture the behavior, as the instantaneous plume structure was found to only weakly depend on the wall-normal distance. Perhaps the most interesting discrepancy with previous work is with the angle of maximum two-point concentration correlation as compared to those measured by \citet{TalluruChauhan-2020}. Our current measurements show a mean angle, $\alpha\approx70^\circ$ (figure~\ref{fig:InclinAngle}), which is different from $\approx 30^\circ$ that was observed by \citet{TalluruChauhan-2020} for similar $x/\delta$, $h_i/\delta$ location (using two-probe measurements and Taylor's hypothesis). 
The stream-wise extent for the correlations in the current work was also found to be smaller. \RevisionText{Though the outer-scaled injection location, $h_i/\delta$ is matched, there are differences in the relative source diameters ($d_s/\delta$) and relative plume sizes ($\sigma_y/\delta)$ between the two efforts which can affect mean vorticity experienced by the plume}. It appears that the boundary layer scaling, $x/\delta$ is not an appropriate scaling parameter to capture the stream-wise variation of the angle \citep{Miller-2005}. A more focused study is required to understand the relevant evolution and scalings of these quantities.

\subsection{On the plume intermittency and breakup}
The plume behavior is highly intermittent in the stream-wise region of interest, with intermittency values, $\gamma\approx60\%$ similar to those observed in \cite{CrimaldiWileyEtAl-2002}. The availability of the spatial information enabled us to explore this behavior as `blobs' of scalar plume as viewed in the plane of measurement. These blobs sustain locally high values of scalar concentration relative to the unconditional mean. Specifically, we looked at the shape, size, mean concentration and inclination of these conditional blobs. The distribution of the concentration and the blob size indicated that \textit{stretching-straining} motions and the plume \textit{breakup} due to large turbulent scales is the primary mode of plume evolution, and that there was very little small-scale scalar mixing. This was true for both injection locations, and can be explained by the relative absence of small-scale turbulence away from the wall compared to the buffer layer \citep{NgMontyEtAl-2011a}. Further, for both injection locations, the distribution of blob parameters did not significantly depend on the wall-normal distance from the injector. This further strengthens the meandering plume hypothesis, and shows that the plume has not had time to significantly adapt to local turbulent conditions at different wall normal locations. Finally, the distribution of the blob inclination angle, $\tilde{\phi}$ and the elongation, $\widetilde{AR}$ show that they are inclined forward (downstream direction) and stretched by around 18\%. These observations bolster the trends observed previously by \citet{TalluruChauhan-2020}. The inclination was slightly higher for the log-region injection compared to the wake region due to the increased shear that the structures encounter as they meander away from the source.

\subsection{On the role of coherent vortices}
\begin{figure}
    \centering
    \includegraphics[width=0.8\textwidth]{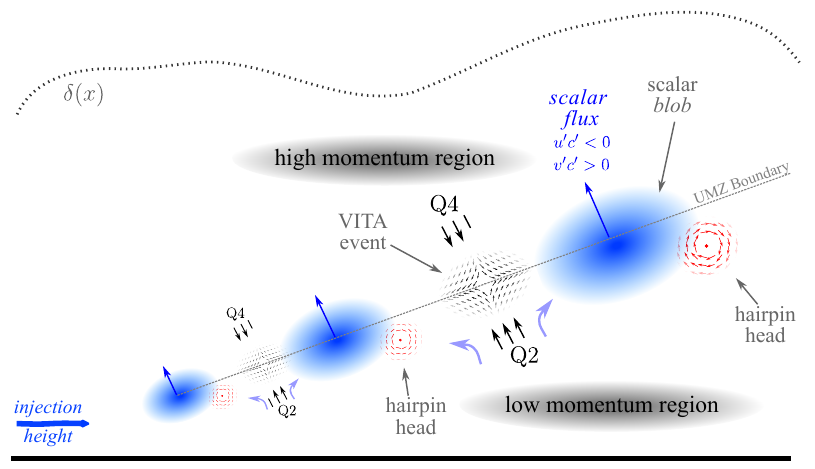}
    \caption{Schematic of meandering of the plume by the coherent vortex packet}
    \label{fig:CVP-paradigm}
\end{figure}
Investigation of scalar blobs relative to the turbulent structure indicated interesting patters. Firstly, and unsurprisingly, it was observed that the swirling motions within the boundary layer are the primary mechanisms by which the plume meanders and breaks up within the boundary layer. The blobs exhibited a correlated arrangement with the swirling strength for both injection locations. This was particularly interesting given the existence of well-known coherent packets of hairpin vortices, that result in large- and very-large scale motions (LSMs and VLSMs). Upon a further investigation using linear stochastic estimations, the following coherent trends in scalar plume arrangement were observed:
\begin{itemize}
    \item Along the line of injection, (i.e. $y \sim h_i$), the scalar blobs were generated via the action of coherent and incoherent swirling motions, and the contributions from the former were not statistically significant to the overall scalar variance. This was true for both injection locations (log- and wake-).
    \item However, plume interactions with a vortex that is a part of a coherent packet shows a distinct footprint along the shear-layer inclined at $\approx 10-15\degree$ (figure~\ref{fig:lse-cvp-log-injection-p1}c). This demonstrated that the packets coherently organize the scalar plume. This is particularly dominant for the log-region injection, and was backed by multiple instantaneous realizations.
    \item More importantly, this coherent mechanism was dominant and statistically significant when the scalar plume meanders away from the injection location significantly, especially for the log-region injection. In other words, when a scalar blob was observed at locations $y \gtrsim h_i + 1.5\sigma_y$, this was almost always associated to the action of a coherent packet. This was evidenced both from the conditional concentration fluctuation at this location and from the conditional vortex events (with their correlated concentration fields). We refer to this as the `\textit{large meander event}'.
    \item The `large meander event' that is coupled with a coherent packet also showed a distinct spatial arrangement in plume structure. Specifically, a strong scalar blob meandering far away from the wall was seen to have correlated, secondary blobs along the shear-layer associated with the secondary vortices of the vortex packet. This was observed only for the log-region injection where the injected plume interacts with the earlier generations of vortices close to the wall. The wake-region injection only interacts with larger, older generation of vortices and do not show this arrangement.
    \item Finally, the region between the scalar blobs was found to sustain, on average, a saddle-like vector topology, commonly known as a VITA event for the coherent vortex packet \citep{AdrianMeinhartEtAl-2000}. \RevisionText{The observed trends are highly consistent with those of \citet{AntoniaFulachier-1989} and \citet{LaskariSaxton-FoxEtAl-2020} who observed the structure of large-scale events.}
\end{itemize}

Based on the above observations, we can propose a mechanistic paradigm (illustrated in figure~\ref{fig:CVP-paradigm}) for the meandering of the scalar plume to add dynamical detail to that presented in \citet{TalluruPhilipEtAl-2018}, \citet{EismaWesterweelEtAl-2021} and \cite{VanderwelTavoularis-2016}: When the plume is injected in the logarithmic region, it interacts with a range of vortices that cause it to meander in the span-wise and wall-normal directions. However, some of these vortices are part of a larger vortex packet that induce a coherent forcing of the plume. The plume appears to respond to this, and the meandering occurs in- and out- of the plane along the inclined shear layer. While this is not a statistically dominant meandering mechanism along the injection line, this appears to be the primary mechanism by which the scalar plume reaches the outer regions of the boundary layer. This has important implications on the \textit{rare, strong-meander} events, and provides a phenomenological explanation for the rare concentration fluctuations that occur far away from the injection location. \RevisionText{Importantly, the current work demonstrates that a scalar from a point source gets transported along the inclined interface of the UMZs over large wall-normal distances (figure~\ref{fig:umz-features}), while also being discretized in a spatially coherent manner by action of the associated coherent vortex packets.}
A Lagrangian analysis of the coherent packet evolution will educate this paradigm more comprehensively. For a plume that is injected in the wake-region, the coherent packet still appears to have a significant role in arranging the scalar into discrete blobs. This is particularly true as the blobs were found coherently placed with vortices. However, this wasn't found to be a dominant contributor of scalar variance at any wall-normal location, at least for the injection location considered in the current work. This is possibly due to the plume's interactions predominantly with older generations of vortices, some of which will be detached from the surface and lack spatial coherence.

In summary, the distribution of strong vortical motions within the boundary layer appears to have a disproportionate influence on the meandering and intermittency characteristics of a passive scalar plume, particularly close to the injection location. Coupled with their role in the sustenance of uniform concentration zones (UCZs) at further downstream locations \citep{EismaWesterweelEtAl-2021} and the correlation between the inclined shear-layers and scalar gradients observed before \citep{LaskariSaxton-FoxEtAl-2020}, the dynamical significance is clearly evident. Given the well established scaling of these structures \citep{MarusicMonty-2019}, it will be interesting to investigate how these translate into their role on passive scalar transport in atmospheric flows.  

\appendix
\section{An alternative heuristic measure of conditional structure}\label{sec:appendix}
We present an alternative and independent perspective of the coherent scalar arrangement that is more intuitive than that presented in section~\ref{sec:coherent-scalar-arrangement}, albeit being less rigorous, less converged, and more subjective. This builds on the observations in instantaneous fields (figures~\ref{fig:instantaneous-coherence} and \ref{fig:umz-features}), conditional-estimate fields (figure~\ref{fig:lse-cvp-log-injection-p1}(a,c)) and the paradigm proposed (figure~\ref{fig:CVP-paradigm}). We ask the question: \textit{given an approximate arrangement assumed between a large meander event and a swirl event, what would a relative conditional flow field be?} For this, `events' satisfying specific conditions were identified as follows (thresholds were based on the illustrative field in figure~\ref{fig:instantaneous-coherence}a):
\begin{itemize}[leftmargin = *, labelsep=*, itemsep=1pt]
    \item \textbf{Scalar event:} a large region [$\hat{A}_{\xi}>\hat{A}_{t}\approx (0.05\delta)^2$] of high concentration [$\xi>0.22$] occurs away from the wall [$\hat{x}, \hat{y}>0.25\delta$]. A total of $\hat{N}_\xi = 2347$ events were observed (out of the 2434 fields captured).
    \item \textbf{Swirl event:} a large region [$\hat{A}_{\lambda}>\hat{A}_{t}\approx (0.05\delta)^2$] of high clockwise swirl [$\lambda_{ci} < -675\delta/U_{inf}$] occus away from the wall [$\hat{N}_\lambda = 11,477$ events in total].
    \item \textbf{Paried events:} Subset of these two events, where they occur simultaneously with the following arrangement, were extracted:
    \begin{itemize}
        \item The swirl event occurs downstream and above the concentration event (marked using area centroids), with a minimum separation of $0.01\delta$ in $x-$ and $y-$ directions.
        \item The swirl event occurs within a maximum distance, $\hat{\Delta}<0.21\delta$
    \end{itemize}
    In occassional cases where multiple swirl events satisfy this proximity criterion, the swirl event nearest to the scalar event is considered. This identified $\hat{N}_P=1205$ pairs in the 2434 fields captured.
\end{itemize}

\begin{figure}
    \centering
    \includegraphics[width=0.3\textwidth]{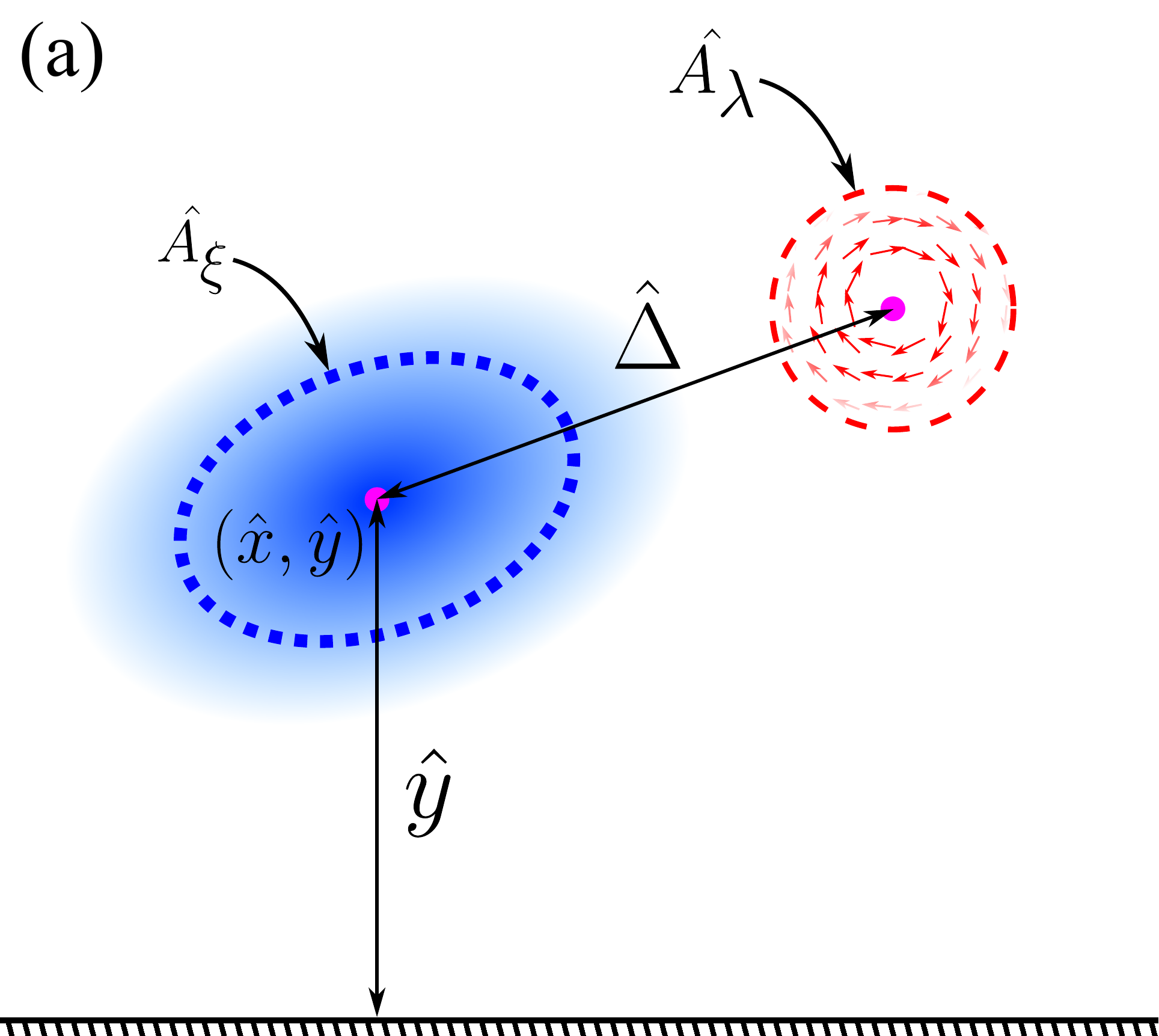}
    \includegraphics[width=0.68\textwidth]{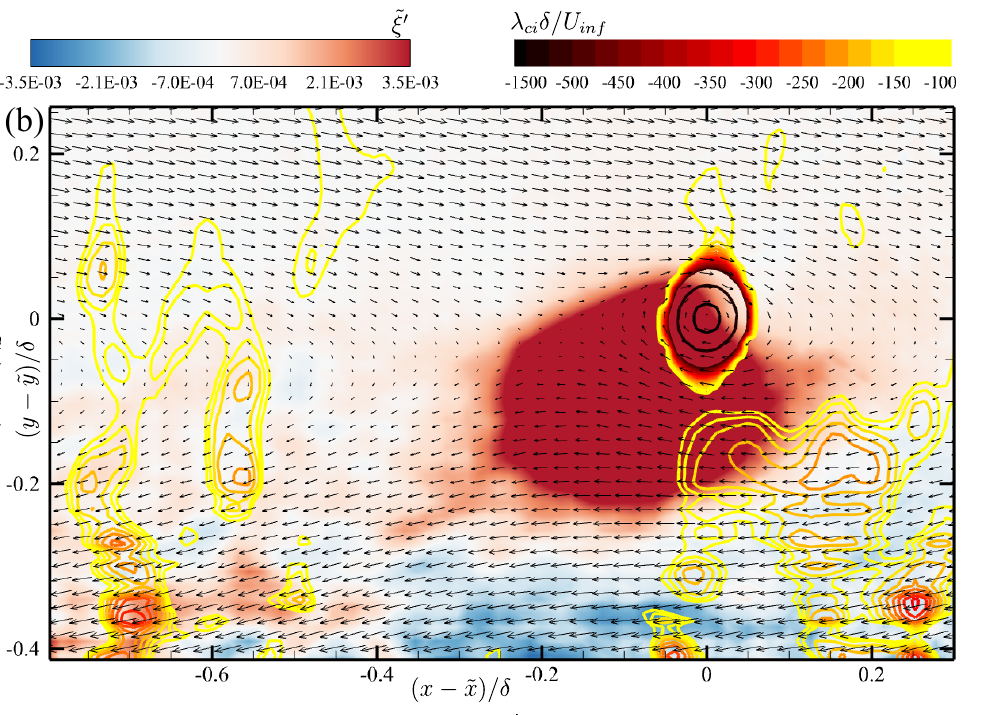}
    \caption{(a)~Schematic representation of the conditional event chosen (b)~Conditional average of concentration fluctuation, velocity and swirl fields of the detected events (log-injection case). Swirl contours are chosen to de-emphasize the conditional event (large swirl) and emphasize the secondary features}
    \label{fig:direct-conditional-averaging}
\end{figure}
This identification and detection method is illustrated in figure~\ref{fig:direct-conditional-averaging}.
The number of detected events implies that there were, on average, 0.495 such pairs per measured field. More crucially, there is 51.3\% probability that a large concentration event is simultaneously accompanied by a large and strong swirl event downstream-above it and within the vicinity chosen.

For all these `\textit{paired events}', the spatial neighborhoods were extracted and conditionaly averaged, to get the relative turbulence field and $\xi'$ field (from the frame of reference of the swirl event). The conditional averaged fields are shown in figure~\ref{fig:direct-conditional-averaging}b. Note that there was no distinction made for paired events occuring at different $x-$ and $y-$ locations (except for the $\hat{y}$-threshold). For this reason, conditional field close to the wall will be averaged across multiple wall-normal locations. This means the near-wall structure observed in these fields will differ to that discussed in section~\ref{sec:coherent-scalar-arrangement}. The conditional structure demonstrates many of the same features identified in figure~\ref{fig:lse-cvp-log-injection-p1}c. The primary swirl-blob coupling is due to the choice of the conditional field. However, a weak secondary peak also appears coupled to a weak swirl, together with an inclined region of negative-$\hat{\xi}'$. The indiscriminate nature of this direct conditional averaging, where multiple packets at different wall-normal distances are averaged, and limited convergence erodes the finer features observed in figures~\ref{fig:lse-cvp-log-injection-p1}c at large distances from reference location. Despite this, the sustainence of primary and secondary blobs, their relative arrangement, and the fact that $\approx 50\%$ of high-concentration blobs occur with swirl event in close vicinity, provides additional statistical evidence to the coupling between the coherent vortex packets and large scalar wall-normal meander.

\backsection[Acknowledgements]{The authors are grateful for valuable insights from Prof. Ronald Adrian on the coherent structures.}

\backsection[Funding]{The research was supported through the start-up grant provided by the Arizona State University}

\backsection[Declaration of interests]{The authors report no conflict of interest.}

\backsection[Data availability statement]{The data that support the findings of this study are available upon request.}

\backsection[Author ORCIDs]{I.E. Wall, https://orcid/0000-0002-5156-4768; G. Pathikonda https://orcid.org/0000-0002-4116-1818}
\backsection[Author contributions]{Isaiah E. Wall contributed by performing the experiments, implementing the diagnostic, and analysis of all results in sections~2-5. Gokul Pathikonda contributed analysis in section~6. Both authors contributed to the experiment conception, writing of the manuscript, and the overall discussions.}
\bibliographystyle{jfm}
\bibliography{2023-ExpFluidsPaper.bib}

\end{document}